\documentclass[useAMS,usenatbib]{mn2e}
\usepackage{epsf}
\usepackage{amssymb}
\usepackage[usenames]{color}
\usepackage{graphicx}
\usepackage{natbib}
\def\plotone#1{\centering \leavevmode
\epsfxsize=\columnwidth \epsfbox{#1}}
\def\plottwo#1#2{\centering \leavevmode
\epsfxsize=.99\columnwidth \epsfbox{#1} \hfil
\epsfxsize=.99\columnwidth \epsfbox{#2}}
\def\plotonebig#1{\centering \leavevmode
\epsfxsize=1.95\columnwidth \epsfbox{#1}}
\def\deg{$^{\circ}~$}

\def\pscm2{\,{\rm phot~s^{-1}~cm^{-2}}}

\title[Positron annihilation spectrum from the Galactic
  Centre]{Positron annihilation spectrum from the Galactic Centre 
  region observed by SPI/INTEGRAL, revisited: annihilation in a cooling ISM? }

\author[Churazov et al.]{E.~Churazov,$^{1,2}$ S.~Sazonov,$^{2}$
S.~Tsygankov,$^{1,2}$ R.~Sunyaev,$^{2,1}$ D.~Varshalovich$^{3}$ \newauthor \\
$^1$ Max-Planck-Institut f\"ur Astrophysik, Karl-Schwarzschild-Strasse 1, 85741
Garching, Germany\\
$^2$ Space Research Institute (IKI), Profsoyuznaya 84/32, Moscow 117997, 
Russia \\
$^3$ Ioffe Physical Techincal Institute, Polytekhnicheskaya 26, St
Petersburg 194021, Russia 
}

\begin{document}

\pagerange{\pageref{firstpage}--\pageref{lastpage}}
\pubyear{2001}

\maketitle

\label{firstpage}
\begin{abstract}
We analyse SPI/{\it INTEGRAL} data on the 511 keV line from the
Galactic Centre, accumulated over $\sim$6 years of observations. We
decompose the X-ray and soft gamma-ray emission of the central part of
the Milky Way into a relatively compact ``Bulge'' and a more extended
``Disk'' components and report their spectral properties.  The Bulge component
shows a prominent 511 keV line and essentially no flux at 1.8 MeV,
while the Disk component on the contrary contains a prominent 1.8 MeV
line and a very weak annihilation line.

We show that the spectral shape of the annihilation radiation (the narrow
511 keV line and the associated othro-positronium continuum) is
surprisingly well described by a model of annihilation of hot positrons 
in a radiatively cooling interstellar medium (ISM). The model assumes
that positrons are initially injected into a hot ($\sim 10^6$~K),
volume filling ISM, which is allowed to freely cool via radiative losses. 
The annihilation time in such a medium is longer than the cooling time for
temperatures higher than a few $10^{4}$~K. Thus, most of the positrons 
annihilate only after the gas has cooled down to $\sim 10^5$~K, 
giving rise to annihilation emission characteristic of a warm, ionized ISM.
\end{abstract}


%

\sloppypar

\section{Introduction}
Although the annihilation line of positrons at 511 keV is the brightest
gamma-ray line in the Galaxy, the origin of the annihilating positrons in
not yet firmly established. First observed with a NaI scintillator
as a $\sim$476~keV line coming from the Galactic Centre (GC) region
\citep{1972ApJ...172L...1J,1973ApJ...184..103J}, it was subsequently
unambiguously identified with a narrow (full width at half maximum,
FWHM$<3.2$ keV) $e^+e^-$ annihilation line using germanium detectors
\citep{1978ApJ...225L..11L}. Since then many balloon flights and
several space missions
\citep[e.g.][]{1991ApJ...375L..13G,1994ApJS...92..387M,1996ApJ...463L..75T,1997ApJ...491..725P} 
have measured the spatial distribution and spectral properties of the
line. 

Since 2003, {\it INTEGRAL} data on the 511 keV line emission became
available. These data show that there is a strong, $\sim
10^{-3} \pscm2$, source of 511 keV photons at the centre of the Milky Way 
with a FWHM of 6--8\deg
\citep[e.g.][]{2003A&A...407L..55J,2005A&A...441..513K,2005MNRAS.357.1377C},
while the 511 keV flux coming from the Galactic plane is much less
constrained by the {\it INTEGRAL} data
\citep[e.g.][]{2005ApJ...621..296T,2008ApJ...679.1315B}. No
significant variability in the narrow 511 keV line has been found in
the SPI data \citep{2010sst}. The spectrum 
of the annihilation radiation from the Galactic Centre region observed
by SPI can be reasonably well described as a combination of a narrow
line ($\sim$2--3 keV Gaussian at 511 keV) and a three-photon
continuum. The flux ratio between the three-photon continuum and the
narrow line suggests that the majority of positrons form positronium
prior to annihilation \citep[e.g.][]{2005MNRAS.357.1377C,2006A&A...445..579J}.

In spite of numerous observations, the origin of the 511 keV emission
from the GC is not established. The problem can be summarised as follows:
\begin{itemize}
\item The morphology of the observed 511 keV line emission does not fit
  the disk-like spatial distribution of such obvious sources of positrons
  as i) massive stars, which produce $\beta^+$ unstable nuclei like
  $^{26}$Al or ii) $\pi^+$'s generated by the interaction of cosmic rays 
  with the interstellar medium (ISM). This favours scenarios with a more
  prominent central excess (``bulge dominated''), including the old
  stellar population (e.g. supernovae Ia), activity of Sgr A$^*$ and the
  annihilation of dark matter particles.
\item The spectrum of the annihilation emission (narrow line width and
  the flux ratio of the line and three-photon continuum) suggests that
  positrons are annihilating in a warm, $\sim 10^4$~K, and slightly
  ionized, $\sim$ a few \%, medium \citep[e.g.][]{2005MNRAS.357.1377C,2006A&A...445..579J}. Such an ISM phase does exist in
  the Galaxy, but its spatial distribution is strongly concentrated
  towards the plane, posing the problem of ``finding'' the right ISM
  phase by a positron.
\end{itemize}

 Recently an attempt to build a self-consistent picture was
  presented by \citet{2009ApJ...698..350H,2009PhRvL.103c1301L}, who assumed
  that the spatial propagation of positrons, produced via $\beta^+$ decay
  of $^{56}$Ni, $^{44}$Ti and $^{26}$Al, is governed by a diffusion process
  with the effective diffusion coefficient different in the bulge and
  the disk of the Galaxy. During diffusion the positrons enter the HII and
  HI envelopes of molecular clouds, in particular those forming a
  ``Tilted Disk''\citep{2007A&A...467..611F} within 1.5 kpc of the
  GC. The model of \citet{2009ApJ...698..350H}, with a reasonable set
  of assumptions, can explain the basic properties of the annihilation
  radiation.

Here we take an alternative route and consider a possibility that a
significant fraction of positrons are born in the hot ISM, which is
eventually able to cool via radiative losses and perhaps via adiabatic
expansion. We show below that with these assumptions the spectral
properties of the 511~keV line can be easily explained. However, to make our
picture self-consistent one needs to model the thermal state of the ISM,
which is beyond the scope of the present paper.

This paper is based on the {\it INTEGRAL} data accumulated over $\sim
6$ years of observations and aims at placing tighter constraints on
the spectral and spatial properties of the annihilation emission. 

\section{Data and background handling}
SPI is a coded mask germanium spectrometer on board {\it INTEGRAL}
\citep{2003A&A...411L...1W}, launched in October 2002 aboard a PROTON
rocket. The instrument consists of 19 individual Ge detectors, has a
field of view of $\sim$30\deg (at zero response), an effective area at
511 keV of $\sim 70$~cm$^2$ and energy resolution of $\sim$2 keV
\citep{2003A&A...411L..63V}. The good energy resolution makes SPI an
appropriate instrument for studying the spectrum of $e^+e^-$
annihilation emission. 

For our analysis we use all data available to us by mid-2009,
including public data, some proprietary data (in particular, proposals
 0420073, 0520071 and parts of 0620059). Prior to actual data analysis, 
all individual observations were screened for periods of very high particle
background. We use the SPI anticoincidence (ACS) shield rate as a main
indicator of high background. Several additional observations were
also omitted from the analysis, e.g. those taken shortly after SPI
annealing procedures \citep{2003A&A...411L..91R}. For our analysis we
used a combination of single and pulse-shape-discriminator (PSD)
events \citep[see][for details]{2003A&A...411L..91R} and treated them in the
same way.

\begin{figure*}
\plotonebig{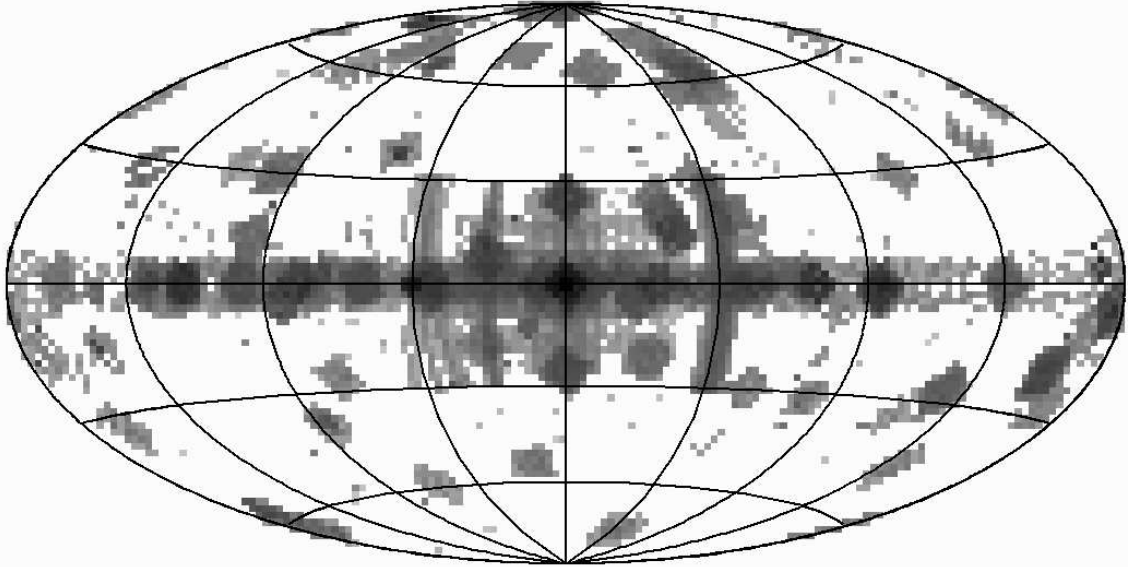}
\caption{Exposure map (in Galactic coordinates) showing all the observations 
  used in the present
  analysis. Off-Galactic-plane observations were used for determining the
  instrument background. For each observation, the effective observing
  time is added to the pixel in the map containing the pointing
  direction of the telescope's axis. The total exposure is about
  70 Ms. Pixels are 2 degrees on a side.
\label{fig:emap}
}
\end{figure*}

\subsection{Energy gain and resolution}
For each detector, a linear relation between the energy and the channel
number was assumed and calibrated (separately for each orbit), using
the observed energies of lines at $\sim$ 198, 438, 584, 882, 1764,
1779, 2223 and 2754 keV \citep[see][for a comprehensive list of SPI
  background lines]{2003A&A...411L.113W}. With this calibration the
RMS deviation of the background 511 keV line energy (revolution based)
is 0.0066 keV while the mean energy of the line is 510.926 keV. There is
thus a small deviation of the line mean energy from the electron rest energy 
(510.999 keV): $\Delta E=0.07$ keV; it can be attributed to the
simplified linear energy/channel relation. This deviation is
comparable to the statistical uncertainty on the line energy (see
\S\ref{sec:spec}) and no attempt was made to correct for this effect.

Since the background 511 keV line (produced by the positrons
annihilating in the body of the detector) is kinematically
broadened, we used the two bracketing lines (at 438 and 584 keV) to
calculate the resolution at 511 keV as $\displaystyle {\rm
  FWHM}_{511}=0.5\times({\rm FWHM}_{438}+{\rm FWHM}_{584})$. The
resulting value is ${\rm FWHM}_{511}=2.175~{\rm keV}$ when averaged
over all observations in the vicinity of the Galactic Centre.

\subsection{Background modeling}
\label{sec:background}
In the background modeling we followed the scheme used in
\cite{2005MNRAS.357.1377C}. Namely, the background count rate
$B(i,E)$ in detector $i$ at energy $E$ is assumed to be proportional to the 
detector saturated (i.e. above 8 MeV) event rate $R_{\rm sat}$ and time $t$:
\begin{eqnarray}
B(i,E)=C(i,E)+\alpha(i,E)\times R_{\rm sat}+\beta(i,E) \times t,
\end{eqnarray}
where $i=1,19$ is the detector number.  The
coefficients $\alpha(i,E)$ and $\beta(i,E)$ of this linear relation
were determined separately for each detector/energy channel using all
the available SPI data, while the constant $C(i,E)$ was estimated using
only the data away from the Galactic plane and away from bright
sources. The whole data set was divided into 14 time intervals (defined by
the annealing periods and the dates of individual detector
failures\footnote{Three out of the 19 SPI detectors had failed by the
  end of 2009.}) 
and the coefficients $\alpha(i,E)$, $\beta(i,E)$ and $C(i,E)$ were determined
separately for each interval. This procedure, although not providing a
perfect description of the background, works reasonably well at all
energies and, given the small number of free parameters, is very
robust. For instance, the relative RMS deviation of the 600--1000 keV
flux (averaged over one revolution) from the count rate predicted by
this model is $\sim 1.3~10^{-3}$.

Along with our reference background model, we made several experiments
by dividing the full data set into different time intervals 
and modifying the selection criteria used for cleaning the data from 
background flares. In these experiments we generated
several ``alternative'' background models. Below we use these
models to test the sensitivity of our results to the details of the 
background modeling.

\section{Imaging}
Imaging with coded mask telescopes often requires additional
assumptions to be made on the sky surface brightness distribution,
e.g. sparsity of compact sources or regular behaviour of diffuse emission. 
This is especially important for the SPI telescope,
which has only $n=19$ independent detectors/pixels. This makes image
reconstruction in a single observation problematic, given that a typical
image reconstruction procedure relies on the smallness of the
parameter $1/n$ with respect to unity. We have chosen not to make 
a priori assumptions about the spatial distribution of sources and restricted 
ourselves to three different types of ``imaging'': (i) light bucket imaging, 
(ii) simple parametric models and (iii) linear decomposition of the sky 
surface brightness into few simple templates.

\subsection{Light bucket mapping}
\label{sec:bucket}
In this simple ``imaging'' procedure, for each observation, the
background subtracted counts in a given energy band are 
summed over all detectors and the resulting signal is added to the pixel of
the map corresponding to the telescope pointing direction. Similarly 
the exposure map is formed by adding the exposure times (see
Fig.~\ref{fig:emap}). The ratio of the counts and the exposure map
gives the count rate image, which is crudely converted into units
of ${\rm phot/s/cm^2}$ using the effective area for an on-axis point
source. Such maps will provide a correct point source flux if the
telescope is pointed directly towards the source. But given the large
size of the SPI FoV ($\sim$30\deg diameter at zero response) and
sparse coverage of the sky, 'light bucket mapping' produces rather
crude maps, while the procedure itself is very simple and robust. We
use light bucket mapping to get a global surface brightness map or to
follow the variations of the surface brightness along and across the Galactic
plane with poor angular resolution. Note that in this mode SPI is
used as a collimated instrument and its mask imaging capabilities are
not exploited. The net flux (difference between the observed count rate and 
predicted background model) crucially depends on the quality of the
background model.

\subsubsection{Scans along the Galactic plane}
\label{sec:longscan}

\begin{figure*}
\plotonebig{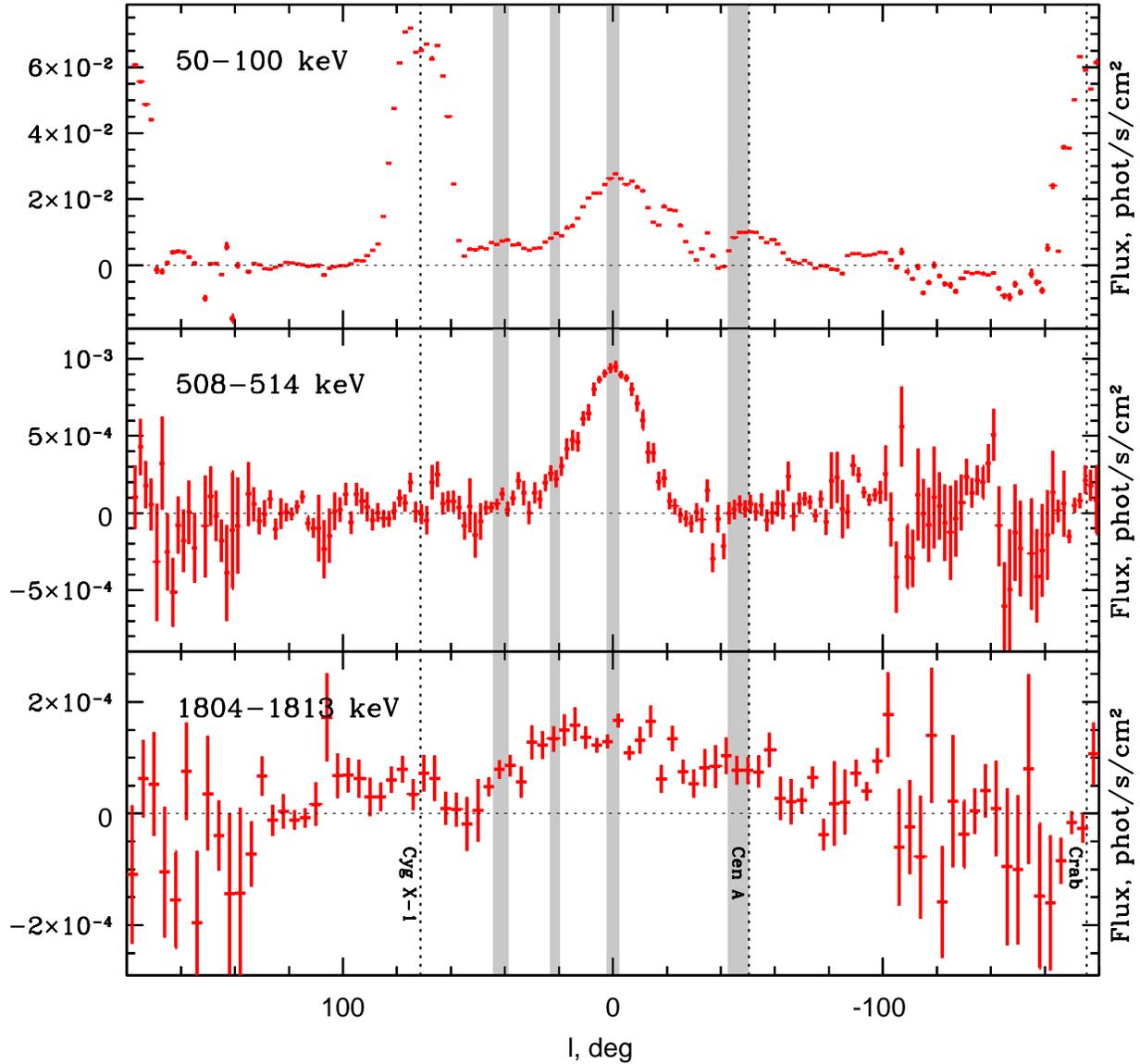}
\caption{Scans (light bucket imaging) along the Galactic plane
  in 3 energy bands. Fluxes are averaged over 2\deg bins over
  $l$ and $\pm 8$\deg over $b$. For the 1804--1813 keV band, the
  width of the bins along $l$ is 4\deg. By construction all points
  are statistically independent (except through the background subtraction).
Vertical grey bars show the positions of the series of scans perpendicular to
the Galactic plane discussed in \S\ref{sec:latscan}. The positions of the
Crab Nebula, Cyg X-1 and Centaurus A are marked with the vertical
dotted lines. 
\label{fig:mslice}
}
\end{figure*}

In Fig.~\ref{fig:mslice} we show slices of the Galaxy along the Galactic 
plane in three energy bands. Each slice has a width of
16\deg perpendicular to the Galactic plane and is centred at
$b=0$\deg. The step along the Galactic plane is 2\deg for the two upper
panels and 4\deg for the bottom panel.

In the upper panel (50--100 keV), three prominent peaks correspond
(from left to right) to Cyg X-1, Galactic Centre and the Crab
Nebula. The width $\sim$30\deg and the complex structure
of the peaks near the maximum (clearly visible for Cyg X-1) are due
to the presence of the mask and partly due to the intrinsic
variability of the sources. The few regions with negative fluxes correspond
to observations where the actual background count rate is higher than
the model-predicted background described in
\S\ref{sec:background}. Note that the 
model background was calculated using a set of ``blank fields'', whose
definition is problematic given the large size of the SPI FoV. The sets of
blank fields are different in each of the 14 time intervals used in the
background modelling (see \S\ref{sec:background}). For this reason,
it is possible that the actual flux from the ``blank fields'' is higher 
than that from some patches of the Galactic place free of strong sources.

The morphology of the slice in the 508--514 keV band\footnote{See
  \citet{2005ApJ...621..296T} for an
  earlier version of a similar plot.}, containing the
511 keV line (Fig. \ref{fig:mslice}, middle panel), is markedly
different. With the 2\deg bins along the Galactic plane, the only
prominent feature is the peak at the Galactic Centre, with
the extent along $l$ roughly similar to the extent of the peak near
Cyg X-1 at low energies. The peak flux is $\sim 10^{-3}~\pscm2$. As
already emphasized, this value is not a precise measure of the true
511 keV flux from the Galactic Centre region, but it should be
accurate to a factor of better than 2 if the source size does not
exceed several degrees. 

No strong asymmetry of the positive and negative longitude wings of
the central peak is apparent in the light bucket profiles (see, however,
\citealt[][who report an excess flux for negative
  longitudes]{2008Natur.451..159W} and \citealt{2008ApJ...679.1315B},
who do not find strong evidence for asymmetry). An
  excess at negative longitudes $l\sim$ -25\deg is visible in the
  light bucket profile in \citet{2005ApJ...621..296T} (their
  Fig. 2). Renormalization of their fluxes to the units used in
  Fig. \ref{fig:mslice} suggests an excess at the level  $\sim
  1.5~10^{-4}~\pscm2$. In our analysis the flux at  $l\sim$ -25\deg is
  essentially consistent with zero. If anything, the
profiles shown in Fig. \ref{fig:mslice} suggest an excess in the
508--514 keV flux at positive longitudes $l\sim$ 25\deg.  However, our
experiments with ``alternative'' background models mentioned in
\S\ref{sec:background} showed that a weak spurious asymmetry
(corresponding to a flux excess/deficit of $\sim 10^{-4}~\pscm2$ at
$|l|\sim$20--25\deg) can appear at either negative or positive
longitudes. We therefore conclude that, with the present knowledge of
the SPI background, light bucket profiles do not
provide compelling evidence for asymmetry in the 508--514 keV
flux along the Galactic plane.

Finally in the bottom panel of Fig. \ref{fig:mslice}, the slice of the
Galactic plane in the 1804--1813 keV band, containing the 1.8 MeV line of
$^{26}$Al, is shown. Unlike the 511 keV emission, there is a broad
peak (much broader than the SPI response to a point source), centred at
the GC. This distribution is qualitatively consistent
with the Comptel \citep{2001ESASP.459...55P} and earlier {\sl INTEGRAL}
results \citep{2009A&A...496..713W}.

\subsubsection{Scans perpendicular to the Galactic plane}
\label{sec:latscan}

\begin{figure*}
\includegraphics[width=1.99\columnwidth,bb=0 360 580
  720,clip]{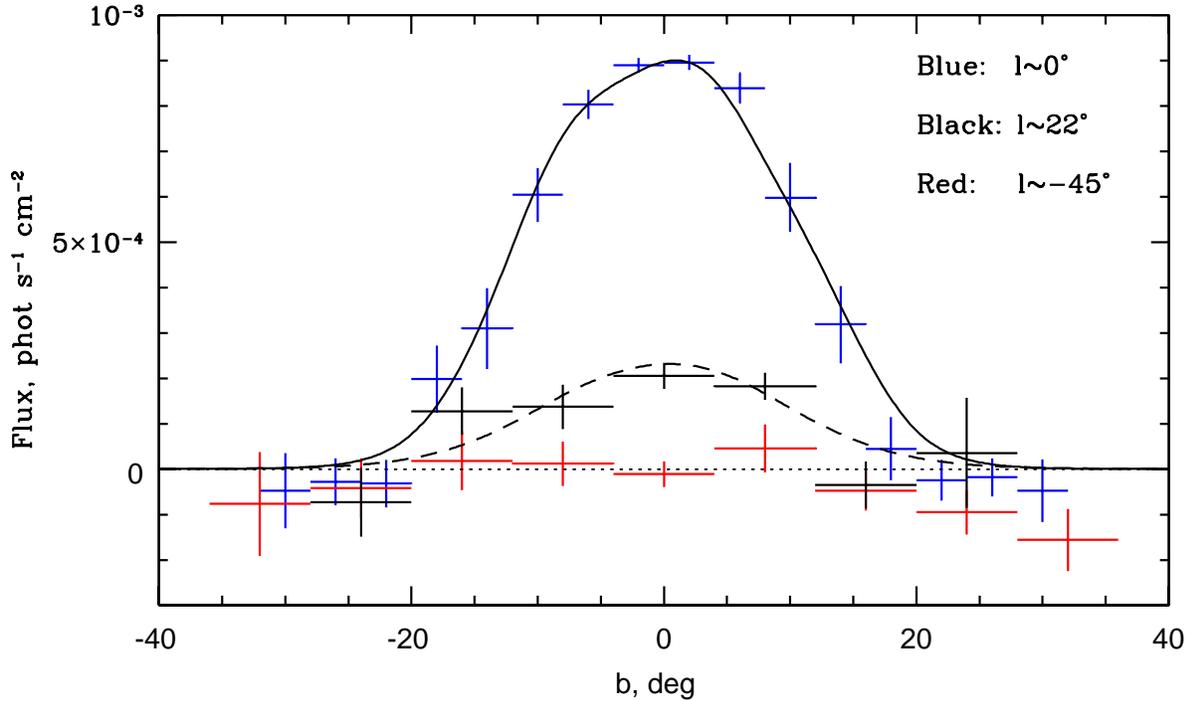}
\caption{ 
 Scans (light bucket imaging) perpendicular the Galactic plane
  at $l\sim$ 0\deg, 22\deg, -45\deg in the 508--514 keV energy band.
  The solid and dashed curves approximately show the expected
  profiles at $l\sim$ 0\deg and 20\deg if the spatial distribution of
  the flux in the 508--514 keV band is adequately described by the
  ``bulge'' component described in \S\ref{sec:templates}.
\label{fig:vslice}
}
\end{figure*}

A similar approach of light bucket mapping can be applied to the 
special scans of the Galactic plane along $b$ performed in 2007, 2008
and 2009. The 
positions of the scans in $l$ are denoted with the gray vertical lines
in Fig.~\ref{fig:mslice}. Three out of four scans were made as a
sequence of pointings starting and ending $\sim 30$\deg away from the plane 
on either side of it. 
The pointing direction was changing in $\sim$2\deg steps. The
remaining ``scan'' at $l=0$ had been done earlier as a set of pointings
directly at the plane and $\sim$25\deg above and below the plane. 
The advantage of making ``fast'' (taking 10--15 hours) scans over a
selected region of the sky is that the quality of the background
modeling can be verified/improved if typical variations of the
detector background occur on longer time scales.

The scans in the 508--514 keV band in the direction perpendicular to
the Galactic plane at $l\sim$ 0\deg, 22\deg, -45\deg are shown in
Fig.~\ref{fig:vslice}. The signal is clearly visible at 
$l\sim$ 22\deg and is close to zero for $l\sim$ -45\deg. To guide the eye,
we have plotted with the black curves the expected light bucket profiles for
the case where the spatial distribution of the flux in the 508--514 keV band
follows the ``Bulge'' model described later in
\S\ref{sec:templates}. The expected profiles were crudely estimated
assuming that the axis of the instrument was moving along $b$ at fixed
$l$, while keeping the position angle fixed. This causes visible
asymmetry in the peak of the $l\sim$ 0\deg profile, reflecting the
particular orientation of the SPI coding mask relative to the
source. In reality, the observed profiles are combinations of
observations with varying position angle 
and varying $l$ and hence they may differ from the simulated profiles
in subtle details. Broadly, the light bucket imaging shows the
overall consistency with a bulge-dominated distribution of the
508--514 keV flux and indicates that the disk emission is weak at
$l\sim$ -45\deg. 

\subsection{Simple parametric models}
\label{sec:par}

We now use a simple function -- two-dimensional Gaussian -- to model the
spatial distribution of the annihilation line emission near the
Galactic Centre. The data collected when the {\sl INTEGRAL} pointing
direction was within 30\deg of the GC were used. Two models were
considered:
\begin{eqnarray}
\begin{array}{ll}
G(l,b)&=F_1\times
e^{-\left\{\frac{l^2\ln 2}{{\rm W_l}^2}+\frac{b^2\ln 2}{{\rm
      W_b}^2}\right\}}, \\
G_D(l,b)&=F_1\times
e^{-\left\{\frac{l^2\ln 2}{{\rm W_l}^2}+\frac{b^2\ln
    2}{{\rm W_b}^2}\right\}}+
 F_2\times
e^{-\left\{\frac{b^2\ln 2}{{\rm W_D}^2}\right\}},
\end{array}
\label{eq:g2}
\end{eqnarray}
where $G(l,b)$ is the surface brightness distribution as a function of
Galactic longitude and latitude, $F$ is the flux and $W_l$, $W_b$ are the 
full widths at half maximum along $l$ and $b$, respectively. The
second model has an additional component that is aimed to account for
emission elongated over the Galactic plane. Since only the data within
30\deg of the GC were
used (30\deg corresponds to the deviation of the SPI axis from the GC), 
the ``infinite'' extent of this component over $l$ means that
the surface brightness of this component does not decrease much at a
distance of $\sim$45\deg from the GC.  

\begin{figure}
\plotone{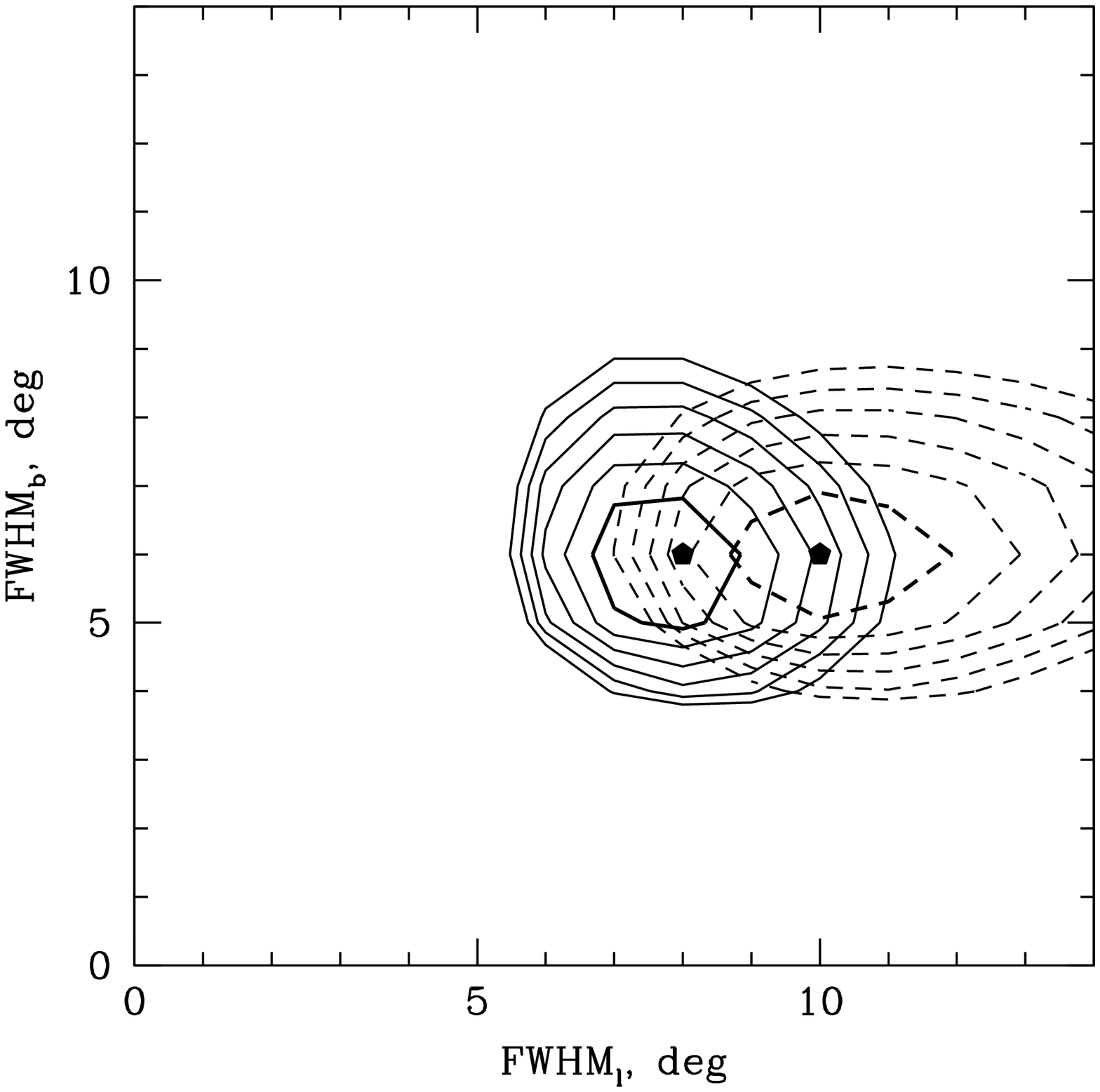}
\caption{Contours of $\chi^2$ as a function of the FWHM along $l$ and
  $b$ when the spatial distribution of the 508--514 keV flux is
  described by a 2D Gaussian. The dashed lines correspond to the pure 2D
  Gaussian model ($G(l,b)$ in eq.~\ref{eq:g2}), while the solid lines
  correspond to the model $G_D(l,b)$ in eq.~\ref{eq:g2}, which includes
  an additional extended component. Black dots mark the positions of the 
  $\chi^2$ minima. Contours are spaced by $\Delta
  \chi^2=3$. The best-fitting values of the width are 10\deg $\times$
  6\deg and 7--8\deg $\times$ 6\deg for the first and second model,
  respectively.  
\label{fig:g2}
}
\end{figure}

To verify the sensitivity of the results to a particular functional
form, we also used an exponential law instead of the Gaussian:
\begin{eqnarray}
\begin{array}{ll}
E(l,b)&=F_1\times
e^{-\sqrt{\frac{l^2}{{\rm W_l}^2}+\frac{b^2}{{\rm W_b}^2}}}, \\
E_D(l,b)&=F_1\times
e^{-\sqrt{\frac{l^2}{{\rm W_l}^2}+\frac{b^2}{{\rm W_b}^2}}}+
 F_2\times
e^{-\frac{|b|}{{\rm W_D}}}.
\end{array}
\label{eq:e2}
\end{eqnarray}

For a given pair of $W_l$ and $W_b$, the model is convolved with the
simulated SPI response \citep{2003A&A...411L..81S} and compared with
the count rate in the 508--514 keV band in individual detectors during
individual observations. The best-fitting value of $F_1$ for the
one-component model or the pair of values $F_1$ and $F_2$ for the
two-component model is calculated. Here and in all the subsequent
analysis, we use simple 
$\chi^2$ statistics. The problems of using the $\chi^2$ criterion for
low photon counting statistics are circumvented following the recipe of
\cite{1996ApJ...471..673C}. Namely, the standard deviation associated
with the count rate in a given spectral, spatial or time bin is
evaluated using the mean count rate averaged over a large number of
``nearby'' similar bins. 

Typical values of $\chi^2$ per degree of freedom for our
models are $\approx 1.01$. Given the large number of degrees of freedom
($\sim 250000=$ number of detectors times the number of observations)
and the very low signal-to-noise ratio of the annihilation signal in
individual observations, this value is not a useful indicator of the
``absolute'' quality of the model. Instead, the change of $\chi^2$ can
be used to compare different models or place constraints on the model
parameters.

The resulting contours of $\chi^2$ for both models (as a function of
the parameters $W_l$ and $W_b$) are plotted in Fig.~\ref{fig:g2}. The
dashed lines correspond to the one-component model (pure 2D Gaussian), 
while the solid lines correspond to the two-component model $G_D(l,b)$ 
(see eq.~\ref{eq:g2}). The black dots mark the positions of $\chi^2$ minima. 
Contours are spaced by $\Delta \chi^2=3$. The best-fitting values of the 
widths are 10\deg $\times$ 6\deg (if the one-component model is used) 
and $\sim$8\deg $\times$ 6\deg (for the two-component model)\footnote{Note 
that we used a grid over FWHM values with a 1\deg step and a 7\deg $\times$
6\deg Gaussian gives almost the same $\chi^2$ as the best-fitting
8\deg $\times$ 6\deg Gaussian.}.

Clearly, for the pure 2D Gaussian model the data suggest a significant
flattening of the distribution towards the plane. If an additional
component, extended along the plane, is included (the two-component model
in eq.~\ref{eq:g2}), then the best-fitting central Gaussian is much more
symmetric in $l$ and $b$. The improvement in $\chi^2$ for the
two-component model compared to the one-component model is $\sim 40$.
The position of the minimum for the two-component model does not depend much
on the width of the extended component over $b$. We tried for the
second component $W_D=2$\deg, $W_D=6$\deg and $W_D=10$\deg as well as an
exponential shape $\displaystyle e^{-\left\{\frac{|b|}{\rm
    W_D}\right\}}$ instead of the Gaussian, and got essentially the
same best-fitting parameters (8\deg $\times$ 6\deg) for the central
Gaussian.

We emphasize here that the presence of the second component does not
necessarily imply that the Galactic disk is "detected", but rather
that a single symmetric Gaussian/exponential component is not a
perfect description of the data.

An exponential shape $\displaystyle e^{-\left\{\frac{|l|}{\rm
    W_l}\right\}-\left\{\frac{|b|}{\rm W_b}\right\}}$ with
$W_l\sim$3\deg and $W_b\sim$2\deg can also be used to describe the
central component (see Table \ref{tab:templates}) instead of a
Gaussian. In fact, the largest improvement in the $\chi^2$ is achieved
when both the central and disk components are described as exponential
functions (Table \ref{tab:templates}).  While the difference in the
$\chi^2$ between various models (e.g. one-component exponential versus
one-component Gaussian) is formally statistically significant, it
corresponds to less than 1\% change in the $\Delta
\chi^2$ value relative to the ``null hypothesis'' of a zero flux in the
508--514 keV band over the entire data set. At this level, a particular value
of $\Delta \chi^2$ for a given model might be sensitive to the
subtle features of the data set. Our experiments with different background
models have indeed shown that the ranking of models in terms of 
$\chi^2$ can slightly vary. For example, the ranking of the Gaussian+Gaussian 
and Gaussian+exponential models in Table \ref{tab:templates} can change
depending on the particular background model. 

Despite the difference in the functional forms, the flux in the central
component does not depend strongly on the model used; it varies 
between $8.4~10^{-4}$ and $9.3~10^{-4} \pscm2$ for all two-components models.

\subsection{Position of the centroid}
Using the one-component exponential model (the first model in
eq.~\ref{eq:e2}) with $W_l=3$\deg and $W_b=2$\deg, we tried to vary the
centroid of the distribution over a few degrees around the GC and
calculated the changes in the $\chi^2$. The best-fitting position is
at $(l,b)=(-0.1^\circ,-0.2^\circ)$ and the minimal $\chi^2$ differs
from the value at $(l,b)=(0,0)$ by 2.3. While this means that formally
the statistics of accumulated data is already sufficient to measure
sub-degree shifts of the centroid, we believe that systematic
uncertainties and the obvious crudeness of the spatial model preclude
any firm conclusion. We can conservatively conclude from 
this exercise that the position of the centroid is consistent 
within $\sim$0.2\deg with the position of the dynamic centre of the Milky Way.

\subsection{Tilted Gaussian}
\label{sec:tilt}
\begin{figure}
\plotone{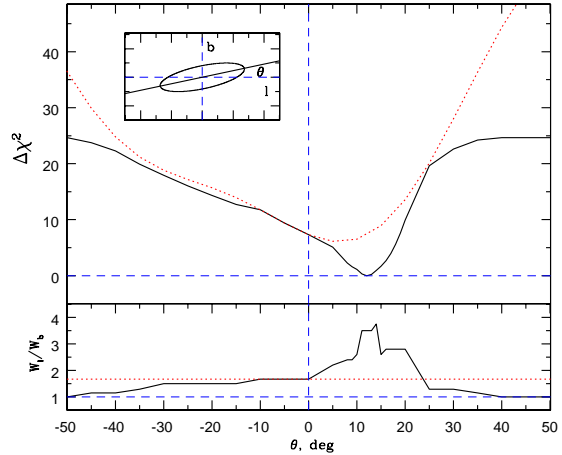}
\caption{ $\chi^2$ as function of the tilt $\theta$ of a two-dimensional
  Gaussian (solid black line). The tilt is measured relative to the Galactic
  plane. Positive values of the tilt correspond to the major axis
  pointing towards positive $b$ at negative $l$. 
   For each value of the tilt the
  best fitting values of the FWHM in two dimensions were identified on
  a crude FWHM grid with steps of $\sim$1\deg. The ratio of the
  FWHMs ($\displaystyle W_l/W_b$) is shown in the bottom panel. The
  minimum of the $\chi^2$ is 
  achieved at $\theta=12$\deg and $W_l=14$\deg, $W_b=4$\deg. The large
  variations of the $\displaystyle W_l/W_b$ ratio are owing to the
  crudeness of the grid and a very shallow minimum in the function
  $\chi^2(W_l/W_b)$ for a given $\theta$.
As $|\theta|$
  increases the Gaussian converges to a symmetric structure
  ($\displaystyle W_l/W_b\approx 1$).
  For comparison, the red dotted line shows the $\chi^2$ as a function of
  the tilt if $W_l$ and $W_b$  are fixed at their best-fitting values
  of 10\deg and   6\deg, respectively, for $\theta=0$.
\label{fig:rot}
}
\end{figure}

 \citet{2009ApJ...698..350H} have suggested that a significant fraction
  of positrons is annihilating in a ``Tilted Disk'' of neutral gas
  inside the central 3 kpc region of the Galaxy. This ``Tilted Disk''
  is one of the components of the \citet{2007A&A...467..611F} model of
  the interstellar gas distribution in the innermost part of the Milky
  Way, based on earlier results of \citet{1980ApJ...236..779L}
  \citep[see also][]{1996IAUS..169..297L}. The Tilted Disk \citep[see
    Fig. 4 in][]{2007A&A...467..611F} in projection to the sky plane
  has an apparent size of $\sim$18\deg by $\sim$5\deg and is tilted by
  $\sim 30$\deg with respect to the Galactic plane. Since the extent
  of the Disk in $l$ and $b$ resembles the dimensions of the 511 keV
  source, \citet{2009ApJ...698..350H} suggest that the positrons reach
  the disk and annihilate there.  

To test further the Tilted Disk model we fit the data with a
two-dimensional Gaussian allowing for rotation of the major axis. The
results are shown in Fig. \ref{fig:rot}. The best-fitting rotation
angle and the widths of the Gaussian in two directions are
$\theta\sim$12\deg and $W_l=14$\deg, $W_b=4$\deg, respectively. The
improvement in $\chi^2$ compared to $\theta=0$\deg (and $W_l=10$\deg,
$W_b=6$\deg) is $\Delta\chi^2\approx 7$. As discussed already, the
values of $\Delta\chi^2$ of the order of a few are 
formally statistically significant if the errors are due to counting
Poisson noise only. The total change of the $\chi^2$ when a
two-dimensional Gaussian is added to the model is $\sim 8000$. Comparing
these numbers, it is clear that even modest systematics in the data
could affect precise derivation of the source spatial
characteristics. Taking the results shown in Fig. \ref{fig:rot} at 
face value, a marginally significant improvement in the fit is possible if the
annihilation region is tilted by $\sim 12$\deg with respect to the
Galactic plane. 

As $|\theta|$ increases beyond 25\deg, the Gaussian converges to an
almost symmetric structure. The tilt of $\theta\sim 30$\deg
corresponds to $\Delta\chi^2\approx 23$ for an almost symmetric
Gaussian with $W_l=9$\deg and $W_b=7$\deg. We therefore conclude that
a version of a the Tilted Disk of \citet{2007A&A...467..611F} is not
particularly favored by SPI data compared to a structure with zero
tilt.  However, moderate values of the title $\lesssim$20\deg are
  allowed by the SPI data.

\subsection{Decomposition using plausible templates}
\label{sec:templates}

\begin{figure}
\plotone{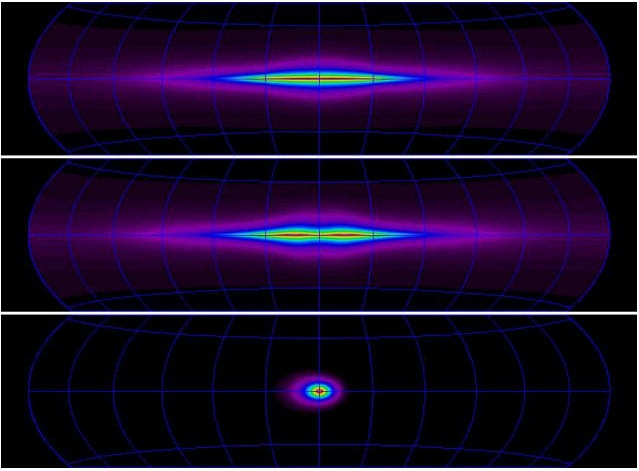}
\caption{The model templates used for fitting the data in the 508--514
  keV band (from top to bottom): Disk, Disk with a Hole, Bulge. The
  projected surface brightness distributions shown are
  based on the 3D models of stellar density distribution adopted from
  \citet{2005A&A...439..107L}. 
\label{fig:mw}
}
\end{figure}

The simple light bucket imaging and parametric fitting of the 508--514
keV flux done in \S\ref{sec:bucket} and \ref{sec:par} suggest that a
reasonable description of the data in the central radian could be
achieved with an almost symmetric Gaussian at the centre and a more
extended component along the plane. The stellar bulge and stellar disk are
the most obvious structural components of the Galaxy that 
qualitatively fit this description. In fact, there are plenty of
disk-like structures (cold gas, massive stars, cosmic ray induced
gamma-ray flux, etc) and very few components that show a prominent
peak at the GC. One of the known centrally-peaked distributions is
that of the NIR light, which is a tracer of the old stellar population of
the Galaxy. For this reason, we decided to restrict our
analysis to three simple templates, corresponding to stellar
components of the Milky Way, adopted from \citet{2005A&A...439..107L} 
and shown in Fig.~\ref{fig:mw}: 
\begin{itemize} \item Disk (eq.~2--5 in L{\'o}pez-Corredoira et al.)
\item Disk with a Hole (eq.~6--7 in L{\'o}pez-Corredoira et al.)
\item Bulge (see \S 3 in L{\'o}pez-Corredoira et al.).
\end{itemize}
This is by no means a comprehensive list of possible templates (see
e.g. \citet{2005A&A...441..513K} for a systematic analysis of
various templates). However, the hope is that these spatial
templates will capture the most basic properties of the 508--514 keV
flux distribution, even if the physical motivation is questionable or the
details of shape are not correct.

We also used slightly modified (truncated) versions of these
templates: for each template we set to zero the surface brightness in
all regions where it is smaller than 10\% of the peak value. The
motivation behind this modification is an attempt to have a template
that has a bulge or disk-type morphology at high surface brightness 
but lacks the extended low surface brightness regions, which might make a
significant contribution to the total flux.

\begin{table*}
\caption{Fitting the 508--514 keV surface brightness with templates and 
parametric models. Quoted $\Delta \chi^2$ values were calculated relative 
to the ``null hypothesis'' of zero flux in the 508--514 keV band over 
the entire data set. The data within 30\deg from the GC were used. 
Quoted fluxes correspond to
the integrated model fluxes (with best-fitting normalization) over a
80\deg $\times$ 80\deg square region around the GC.
\label{tab:templates}
}
\begin{tabular}{lllcc}
\hline 
\multicolumn{2}{c}{Spatial template}  &$-\Delta \chi^2$ & $F_{1}$ &
$F_{2}$ \\
Component 1 & Component 2 & & $\pscm2$ & $\pscm2$  \\ 
\hline
\multicolumn{5}{c}{Templates described in \S\ref{sec:templates} with the surface brightness truncated
  at $10^{-5}$ of maximum} \\
\hline
Disk & - & 6837.7 & $(2.20\pm0.03)~10^{-3}$ & - \\
Disk/Hole & - & 6660.3 & $(2.22\pm0.03)~10^{-3}$ & - \\
Bulge & - & 8252.4 & $(1.12\pm0.01)~10^{-3}$ & - \\
Bulge & Disk & 8252.5& $(1.11\pm0.03)~10^{-3}$ & $(2.4\pm6.4)~10^{-5}$ \\
Bulge & Disk/Hole & 8252.5 & $(1.11\pm0.03)~10^{-3}$ & $(2.0\pm6.2)~10^{-5}$ \\
\hline
\multicolumn{5}{c}{Templates with the surface brightness truncated
  at $10^{-1}$ of maximum } \\
\hline
Disk & - & 6964.0 & $(1.83\pm0.02)~10^{-3}$ & - \\
Disk/Hole & - & 6758.9 & $(1.85\pm0.02)~10^{-3}$ & - \\
Bulge & - & 8229.1 & $(0.97\pm0.01)~10^{-3}$ & - \\
Bulge$^{*}$ & Disk & 8252.9 & $(0.87\pm0.02)~10^{-3}$ & $(2.4\pm0.5)~10^{-4}$ \\
Bulge & Disk/Hole & 8253.0 & $(0.87\pm0.02)~10^{-3}$ &
$(2.3\pm0.5)~10^{-4}$ \\
\hline
\multicolumn{5}{c}{$G(l,b)$ and $G_D(l,b)$ models from
  eq.~\ref{eq:g2} and \ref{eq:e2}} \\
\hline
Gaussian $W_l=$10\deg, $W_b=$6\deg & - & 8223.4 & $(0.96\pm0.01)~10^{-3}$ & - \\
Gaussian $W_l=$8\deg, $W_b=$6\deg & Gaussian, $W_D$=2\deg & 8260.3 &
$(0.84\pm0.02)~10^{-3}$ &  $(2.9\pm0.5)~10^{-4}$ \\
Gaussian $W_l=$8\deg, $W_b=$6\deg & Gaussian, $W_D$=6\deg & 8262.8 &
$(0.84\pm0.02)~10^{-3}$ &  $(3.1\pm0.5)~10^{-4}$ \\
Gaussian $W_l=$8\deg, $W_b=$6\deg & Gaussian, $W_D$=10\deg & 8263.7 &
$(0.84\pm0.02)~10^{-3}$ &  $(3.3\pm0.5)~10^{-4}$ \\
\hline
Gaussian $W_l=$8\deg, $W_b=$6\deg & Exponential, $W_D$=2\deg & 8262.6 &
$(0.84\pm0.02)~10^{-3}$ &  $(3.1\pm0.5)~10^{-4}$ \\
Gaussian $W_l=$8\deg, $W_b=$6\deg & Exponential, $W_D$=6\deg & 8262.6 &
$(0.84\pm0.02)~10^{-3}$ &  $(3.7\pm0.6)~10^{-4}$ \\
\hline
Exponential $W_l=$3\deg, $W_b=$2\deg & - & 8265.4 & $(1.01\pm0.01)~10^{-3}$ & - \\
Exponential $W_l=$3\deg, $W_b=$2\deg & Gaussian, $W_D$=2\deg & 8284.1
& $(0.93\pm0.02)~10^{-3}$ & $(2.1\pm0.5)~10^{-4}$ \\
\hline
\multicolumn{5}{l}{$^*$ - The Bulge component of the Disk+Bulge model is used in the
spectral analysis in \S\ref{sec:spec}.}
\end{tabular}
\end{table*}

 For a given set of templates, the surface brightness for each
  template is convolved with the simulated SPI response
  \citep{2003A&A...411L..81S} yielding an expected count rate in the
  508--514 keV band in individual detectors during individual
  observations. The best-fitting normalizations of the templates are
  then calculated in order to minimize the $\chi^2$ deviation between
  the raw data and the model.

The $\chi^2$ values for the various spatial models are given in
Table \ref{tab:templates}. The fluxes quoted in Table~\ref{tab:templates} 
are the integrated model fluxes (with best-fitting
normalization) over a square 80\deg $\times$ 80\deg region around
the GC. The choice of the region for the calculation of the integrated
model flux is rather arbitrary. The 80\deg $\times$ 80\deg region is
not uniformly covered by observations, and areas with the smallest
errors (largest exposures) dominate in the determination of the best-fitting
normalization. On the other hand, underexposed (or even not observed at
all) areas can still provide a significant contribution to the total
integrated flux. An example of a clear overestimation of the flux due
to this effect is seen for the pure ``Disk'' models in
Table~\ref{tab:templates}. 
These models have poor $\chi^2$, but predict a large integrated flux, 
since their normalization is largely set by the
innermost bright region of the Galaxy. This becomes an especially severe
problem when dealing with fluxes integrated over very large areas
(e.g. the flux from the entire Galactic disk). For this reason, we
quote fluxes integrated over the 80\deg $\times$ 80\deg region rather
than from the whole Milky Way.

Table~\ref{tab:templates} strongly suggests that the spatial
distribution of the 508--514 keV flux in the inner region of the Galaxy
is more extended along the Galactic plane than perpendicular to it. 
This is clear from the parameters of the single-component
parametric models (see also Fig.~\ref{fig:g2}) and from the
comparison of the $\chi^2$ values for the one-component and
two-component (e.g. two Gaussians) models.

Among the one-component models, the exponential model has a slightly
better $\chi^2$ than the Gaussian model or the Bulge
model. For the two-component models, the exponential Bulge +
Gaussian Disk has the smallest $\chi^2$ among the set of models
considered here. It is clear that i) the true distribution of the 508--514 keV
flux is likely more complicated than any of the models used and ii)
the relatively modest (although statistically significant) changes in the
$\chi^2$ among different models (containing the Bulge component)
suggest that the basic properties of the distribution are captured by our
models. 

The uncertainty in the choice of the spatial model directly translates
into the uncertainty of the 508--514 keV flux. In particular, the total
flux of the Bulge component varies among all two-component models
(both parametric and based on templates) from
$0.84~10^{-3}$ to $1.11~10^{-3} ~ \pscm2$. At the same time, the total
Bulge + Disk flux (integrated over the 80\deg $\times$ 80\deg region)
varies from $1.10~10^{-3}$ to $1.14~10^{-3} ~ \pscm2$.

Since some of the spatial templates used in Table~\ref{tab:templates} 
are projections of the 3D models (based on stellar
distribution), it is easy to recalculate the observed flux $F$ into the
total luminosity $L$ of the corresponding component as $\displaystyle
L=4\pi D^2\times F$. Here $D$ depends on the 3D distribution of the
volume emissivity and on the region used to calculate the flux
$F$. The values of $D$ are given in Table \ref{tab:reff}. Using
the effective distances from Table \ref{tab:reff} for the templates
truncated at 10\% of the maximal surface brightness, the total
luminosity of the Bulge component is $\sim 5.0~10^{42}~{\rm
  phot~s^{-1}}$, while the total Disk luminosity is $\sim
2.3~10^{42}~{\rm phot~s^{-1}}$. As noted above, the 
resulting luminosities rely on the model 3D distribution and for this
reason they are model dependent. 

\begin{table}
\caption{Effective distance $D$ needed to recalculate the observed flux from
  the 80\deg $\times$ 80\deg region around the GC into the luminosity for
  different spatial components. Two columns correspond to different
  cutoffs in the surface brightness distribution of each component
  (relative to the maximal surface brightness of this component). 
\label{tab:reff}
}
\begin{tabular}{lcc}
\hline 
Spatial template & \multicolumn{2}{c}{Effective distance, kpc}  \\
& cutoff=$10^{-5}$ & cutoff=$0.1$ \\ 
\hline
Disk & 7.74 & 8.98\\
Disk/Hole & 7.79 & 8.99\\
Bulge & 5.89 & 6.93\\
\hline
\end{tabular}
\end{table}

\section{$^{26}$A\lowercase{l} decay}
\label{sec:al26}
$^{26}$Al is one of the obvious and accountable sources of the 
positrons in the Galaxy \citep[see e.g.][]{2008NewAR..52..440D}. 
This radioactive isotope with the half-life time of $7.17~10^5$ years 
is produced by massive stars, which primarily occupy the disk of the 
Galaxy. The positrons are produced in 81.7\% of decays, with the mean 
energy of 543 keV; in 99.8\% a 1.809 MeV photon is
emitted\footnote{http://www.nndc.bnl.gov/mird/.}. Assuming that 
the fraction of positrons forming positronium is $f_{\rm ps}$ (see
\S\ref{sec:spec}), the fluxes in the 511 keV and 1.8 MeV lines per 1
positron produced via $^{26}$Al decay are: 
\begin{eqnarray}
\begin{array}{ll}
F_{511}&=0.817 \left [ (1-f_{\rm ps} )+\frac{1}{4}f_{\rm ps}\right ]
\times 2=1.63 \left [ 1-\frac{3}{4}f_{\rm ps} \right],\\
F_{1.8}&=0.998.
\end{array}
\label{eq:1.8}
\end{eqnarray}
Here we assume that i) the fraction of para-positronium is $1/4$ and ii)
all positrons annihilating without formation of positronium (fraction
$1-f_{\rm ps}$) produce a narrow 511 keV line. The factor $2$ in the above
expression accounts for the 2 photons produced in two-photon
annihilation. Therefore, $\displaystyle F_{511}=0.409, 0.471$ and $1.637
\times F_{1.8}$, for $f_{\rm ps}=$1, 0.95 and 0, respectively (see
\S\ref{sec:spec} for details).

\begin{figure*}
\includegraphics[width=1.9\columnwidth,bb=0 360 580 740,clip]{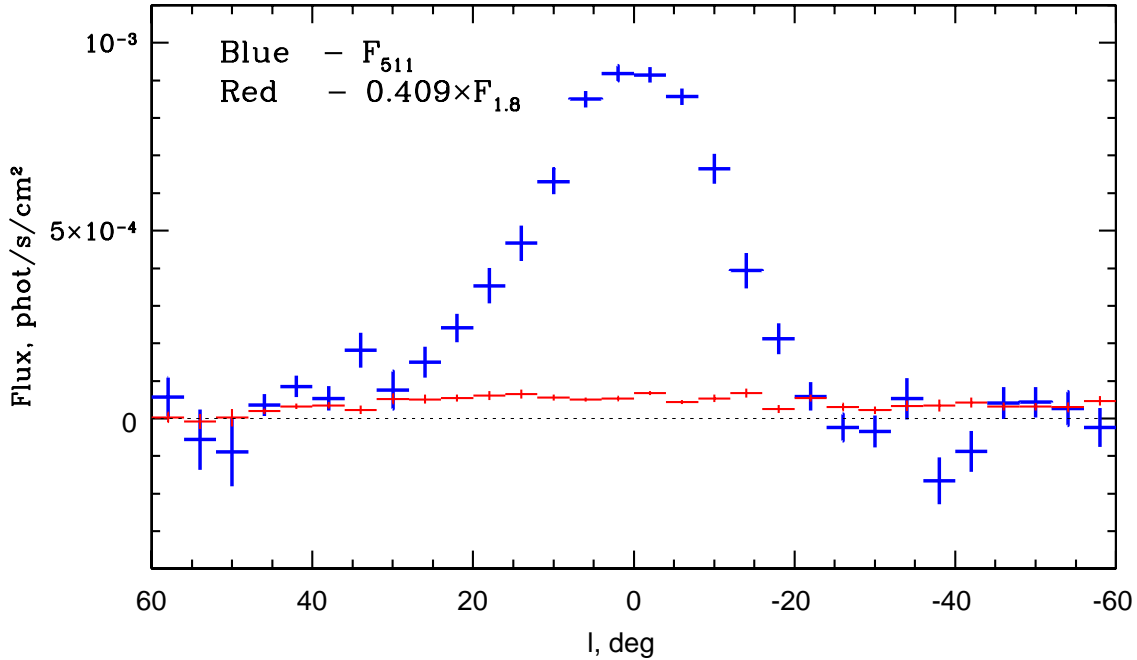}
\caption{Expected contribution of positrons produced by $^{26}$Al
  decay to the annihilation line flux in the vicinity of the Galactic
  Centre. The light-bucket technique was used (see
  \S\ref{sec:bucket}). The slice has a width of 16\deg perpendicular
  to the Galactic plane and is centred at $b=0$\deg. The step along the
  Galactic plane is 4\deg. The measured flux in the 508--514 keV band is shown
  with blue crosses; the red crosses show the flux in the 1804--1813 keV band
  scaled by a factor of 0.409, which is the
  expected 511 keV line flux arising from $^{26}$Al decay under
  the assumption that 100\% of positrons annihilate through the
  formation of positronium. It will be observed later
  (\S\ref{sec:spec}) that this assumption is supported by spectral
  data. Clearly the 1804--1813 keV flux arising from the Galactic
  Centre region is far too low to explain the observed 511 keV line by the
  local $^{26}$Al decay.
\label{fig:alu}
}
\end{figure*}

It is interesting to compare this relation with the fluxes
obtained from the light-bucket images in the 508--514 keV and 1805--1813
keV energy bands, which should contain most of the 511 keV and 1809
keV line fluxes, unless the lines are strongly broadened. Shown in
Fig.~\ref{fig:alu} is a light-bucket longitude scan of the
Galactic plane over the region $l=\pm60$\deg. The blue thick crosses show the
flux in the 508--514 keV band, while the thin red crosses show the
observed flux in the 1805--1813 keV band scaled by a factor
of $0.409$. These red crosses correspond to the expected 511 keV line 
flux arising from $^{26}$Al decay under the assumption that 100\% of 
positrons annihilate through the formation of positronium. It will be 
observed later (\S\ref{sec:spec}) that this assumption is supported by spectral
data. Clearly, the high 511 keV line flux from the GC region cannot be
explained by $^{26}$Al decay \citep[unless the positrons produced in
  the disk are somehow transported to the central region; see
  e.g.][]{2006A&A...449..869P}.  

\begin{figure}
\plotone{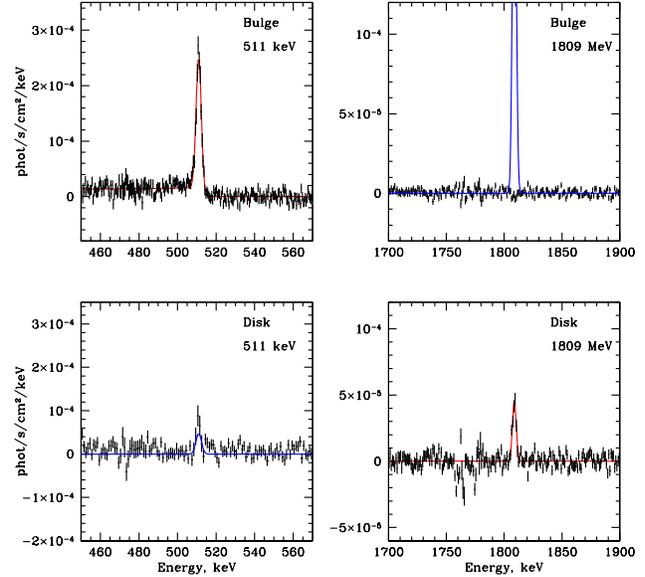}
\caption{Spectra of the Bulge (top row) and Disk (bottom row) near
  511 keV and 1809 keV, obtained using the decomposition of the data
  with the two-component Disk+Bulge model. The red line in the top-left 
  panel shows the best-fitting model (Gaussian at 511 keV and ortho-positronium
  continuum; see \S\ref{sec:spec} below for details). The flux in the
  511 keV line in this model is $F_{511}=0.84~10^{-3}\pscm2$. The blue 
  curve in the top-right panel shows a Gaussian line at 1.809 MeV.  
  The flux in the Gaussian was 
  calculated as $\displaystyle
  F_{1.8}=F_{511}/0.409$, corresponding to the assumption that
  all positrons are produced by $^{26}$Al decay and the fraction of
  annihilations via positronium is $f_{\rm ps}=1.0$ (see eq.~\ref{eq:1.8}
  and \S\ref{sec:spec}). 
The bottom row shows the spectrum of the Disk component. In contrast to
the Bulge spectrum, the 1.8 MeV line is very prominent in the
spectrum (bottom-right panel). The red line shows the best-fitting
Gaussian at 1.809 MeV, with flux $F_{1.8}=4.1~10^{-4}\pscm2$.  The blue
line in the bottom-left panel now shows a 511 keV line with flux
$\displaystyle F_{511}=0.409\times F_{1.8}$. 
\label{fig:aldb}
}
\end{figure}

One can use the same set of templates as in \S\ref{sec:templates} and
repeat the analysis for a set of energies to obtain the spectrum
associated with each spatial component. We did this using the Disk+Bulge
model, which was applied to all observations with the {\it INTEGRAL}
pointing direction within 30\deg from GC. Since we do not explicitly
account for compact sources in this procedure, the low energy parts of
the spectra obtained may not be reliable. However, near the 511 keV and
1.8 MeV lines the contribution of individual sources should not be
crucial. Nevertheless, this simplifying assumption should be kept in
mind when interpreting the resulting spectra, shown in Fig.~\ref{fig:aldb}.  

The upper row in Fig.~\ref{fig:aldb} shows the spectrum of the Bulge component
in the vicinity of the 511 keV and 1.8 MeV lines. The red line in
the top-left panel shows the best-fitting model (Gaussian at 511 keV
and ortho-positronium continuum; see \S\ref{sec:spec} below for
details). The flux in the 511 keV line in this model
is $F_{511}=(0.84\pm0.03)~10^{-3} \pscm2$. Note that this flux was
obtained from the spectral analysis and it differs slightly from the
``Bulge'' flux obtained from the analysis of the count rate in the 508--514
keV band, quoted in Table \ref{tab:templates}. The value $F_{511}$
was then recalculated into the expected flux in the 1.8 MeV line using 
the relation $F_{1.8}=F_{511}/0.409$, corresponding 
to the assumption that all positrons are produced by $^{26}$Al decay and the
fraction of annihilations via positronium is 100\% (see
eq.~\ref{eq:1.8} and \S\ref{sec:spec}). A
line with the resulting flux $\displaystyle F_{1.8}$ is plotted in the 
top-right panel as a Gaussian line centered at 1.809 MeV. Clearly such a strong line at
1.8 MeV is in stark contrast to the data, which show no evidence for
1.8 MeV emission in the Bulge spectrum. 

The bottom row in Fig.~\ref{fig:aldb} shows the Disk
component spectrum in the vicinity of the 511 keV and 1.8 MeV lines. In
contrast to the Bulge spectrum, the 1.8 MeV line is now very prominent in the
data (bottom-right panel). The red line shows the best-fitting
Gaussian at 1.809 MeV with flux $F_{1.8}=(4.1\pm0.5)~10^{-4} \pscm2$. 
The blue line in the bottom-left figure now shows a 511 keV line with flux
$\displaystyle F_{511}=0.409\times F_{1.8}$. The data do show the
presence of a 511 keV line in the Disk component, although our
experiments with the alternative background models and various spatial
patterns demonstrate that the parameters of the line near 511 keV are not
determined reliably. 

Thus, Fig.~\ref{fig:aldb} suggests that with the two-component spatial
Bulge+Disk model: 
\begin{enumerate}
\item The strong 511 keV line is confined to the Bulge component, with
  flux $0.84-0.93~10^{-3}~\pscm2$ and the corresponding 511 keV
  line production rate of $\sim 0.5~10^{43}~{\rm phot~s^{-1}}$ (equivalent
  to an annihilation rate of $\sim 10^{43}$ positrons per second if
    the positronium fraction is 100\%).
\item The 1.8 MeV line is on the contrary confined to the Disk
  component. The flux (model flux integrated over a
80\deg $\times$ 80\deg square region around the GC) is
$4.1~10^{-4}~\pscm2$, corresponding to an $^{26}$Al mass of $\sim
 2.8~M_\odot$ within the same region \citep[see also][]{2009A&A...496..713W}. 
\item An upper limit on the ratio of the 1.8 MeV and 511 keV line
  fluxes in the Bulge component is $\displaystyle
  F_{1.8}/F_{511}<0.027~~(2\sigma)$, which 
  is two orders of magnitude lower than the expectation for the positrons 
  being produced by $^{26}$Al decay and $f_{\rm ps}\sim 1$ 
  (even if $f_{\rm ps}=0$, the ratio is still 
  $\sim$20 times lower than the expectation).
\item The ratio of the 1.8 MeV and 511 keV line fluxes in the Disk
  component $\displaystyle F_{511}/F_{1.8}=0.56\pm 0.2$ is
  consistent with $^{26}$Al decay producing the positrons, 
  although the large uncertainty in the 511 keV flux in the Disk 
  component precludes a firmer conclusion.  
\end{enumerate}

\begin{figure*}
\includegraphics[width=1.9\columnwidth,bb=0 350 582 680,clip]{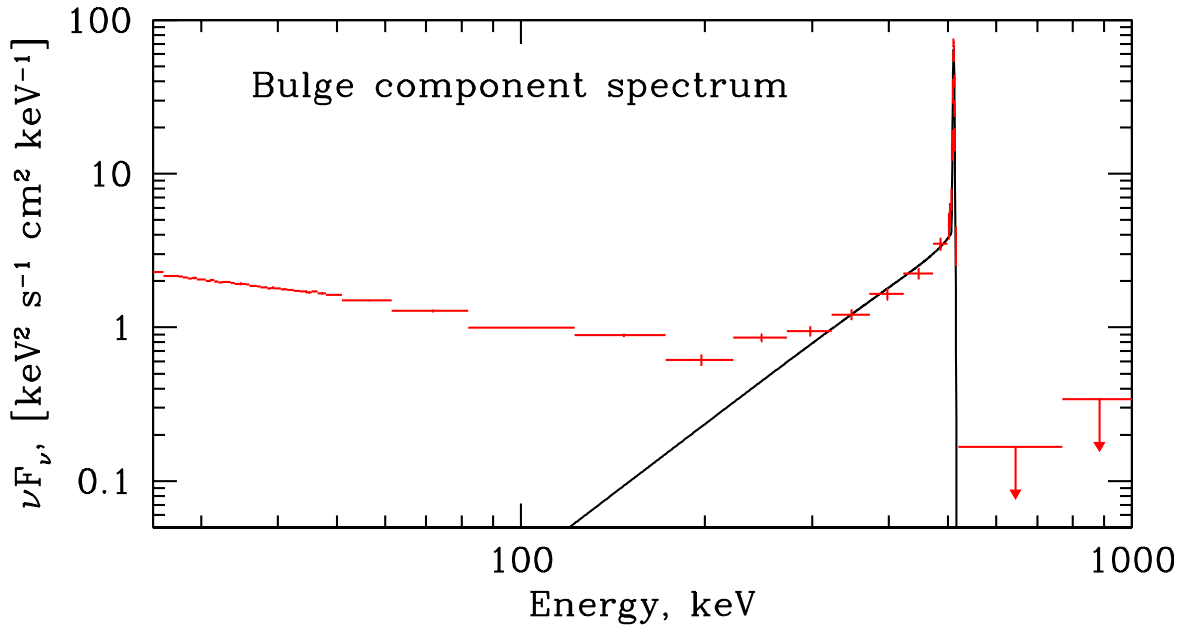}
\caption{Broad 30--1000 keV spectrum of the Bulge component of the
  two-component spatial model described in \S\ref{sec:templates}. The black
  line shows the spectral model consisting of a Gaussian line at
  511 keV and the associated ortho-positronium continuum, 
  with the parameters given in Table \ref{tab:spec}.
\label{fig:broad}
}
\end{figure*}

\section{Spectra}
\label{sec:spec}

The emission seen by SPI from the central radian of the Galaxy
contains contributions of many spatially and physically different
components, e.g. bright compact sources, unresolved weak sources,
emission coming from the disk, etc. The clean ``decoupling'' of 511
keV and 1.8 MeV emission into the Bulge and Disk components made in
\S\ref{sec:al26} strongly suggests that analyzing the spectrum
associated with the Bulge component is advantageous for understanding
the nature of 511 keV emission. Indeed, the Disk component might absorb
some physically unrelated emission (like the 1.8 MeV line or
the continuum at hundreds keV, which may be attributed to the 
interaction of cosmic rays with the ISM), and the remaining Bulge 
component spectrum should be ``cleaner''. The main disadvantage of this 
approach is that the
significance of all features goes down once two spatial components
with free normalizations are allowed (see e.g. Table
\ref{tab:templates} -- the error on the 508--514 keV flux from the Bulge
is twice as large in the Bulge+Disk model as in the pure Bulge
model). In spite of the increased uncertainties, we choose to apply
the subsequent spectral analysis to the Bulge component of the
Bulge+Disk model. The broad-band spectrum of this component is shown
in Fig.~\ref{fig:broad}. 
In the Bulge component spectrum, the flux at energies above 511
  keV is consistent with zero, suggesting that a broad band continuum
  does not contribute much around 500 keV. At energies lower than
  $\sim$300 keV, a broad band continuum is clearly visible. We know
  many compact sources that contribute to this continuum,
  e.g. 1E1740.7$-$2942 or GRS1758$-$258. They have hard spectra around 100
  keV, but decline strongly at higher energies, making a small
  contribution at $\sim$500 keV.

\subsection{In-flight annihilation}

As a rule, the positrons produced by any plausible physical process are
born ``hot'', i.e. with kinetic energy of order of or higher than the
rest energy. For positrons with energy $E_{\rm kin}\le m_ec^2$, 
the rate of Coulomb and ionization energy losses dominates over 
the annihilation rate and the positrons first slow down before 
they have a chance to annihilate. 

\begin{figure}
\plotone{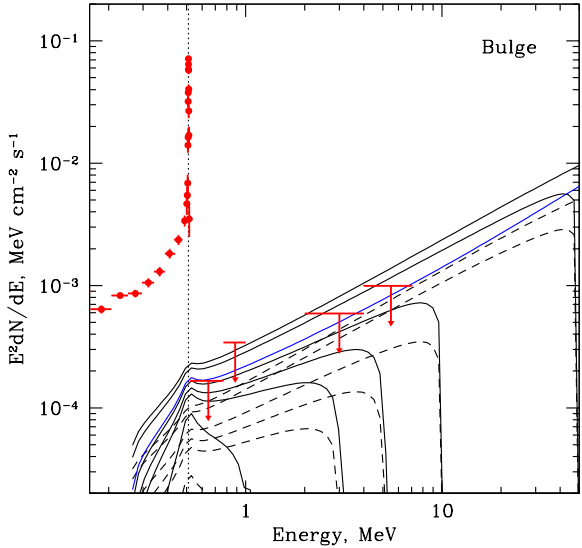}
\caption{Expected in-flight annihilation spectrum for a neutral (solid)
  and ionized (dashed) medium, normalized to the 511 keV line flux
  observed in the Bulge component of the Bulge+Disk model. The initial
  energy of positrons is 1, 3, 5, 10, 50 and 100 MeV. The upper
  limits are 1$\sigma$ values. Both the in-flight annihilation and 
  internal bremsstrahlung components are included in the shown
  spectra, calculated following
  \citet{1981PhLB...99..301A,2006PhRvL..97g1102B}. The blue solid line
  shows the pure in-flight annihilation spectrum  
  (without internal bremsstrahlung) for 100 MeV positrons in a neutral ISM.
\label{fig:ia}
}
\end{figure}

In some scenarios (e.g. dark matter annihilation or production of
positrons in the vicinity of relativistic compact objects), the
initial energy of the positrons can be much higher than $m_ec^2$.  As
discussed by
\citet{1981SvAL....7..395A,1981PhLB...99..301A,2006PhRvL..97g1102B,2006PhRvD..74f3514S}, 
if $E_{\rm kin}\gg m_ec^2$, the energy losses and
annihilation rates differ by only a factor of $\sim$10 and a
substantial fraction of 
high energy positrons annihilate during the slow-down process. These
high energy positrons will form a broad feature at a mean energy $\sim
m_ec^2+E_{\rm kin}/2$, while the slowed down positrons will annihilate at
much lower energies and will power a narrow 511 keV line. Thus, a high
energy (above 511 keV) component is expected to be present in the
spectrum, with the relative intensity with respect to the narrow 511
keV line depending on the initial positron energy $E_{\rm kin}$ and
the ionization state of the interstellar medium. The latter controls the
contribution of the ionization and Coulomb losses to the total
energy loss rate. 

Shown in Fig.~\ref{fig:ia} is the observed spectrum
of the Bulge component and the expected in-flight annihilation
spectrum\footnote{The expected spectrum in addition to the in-flight
  annihilation includes the internal bremsstrahlung
  component \citep{2005PhRvL..94q1301B,2006PhRvL..97g1102B}.} for a
neutral (solid) and ionized (dashed) medium. The difference in
normalization is due to the larger energy losses in the ionized medium,
which increases the fraction of the slowed down positrons at the expense
of in-flight annihilation. The fraction of slowed down positrons
annihilating via positronium formation was set to $f_{\rm ps}=0.97$.  The
spectra shown correspond to initial positron energies of 1, 3, 5, 10,
50 and 100 MeV. No significant flux above 511 keV is observed by SPI in
the Bulge component (for the two-component model described in
\S\ref{sec:templates}), and in Fig.~\ref{fig:ia} we show the corresponding
2$\sigma$ upper limits. Evidently, the SPI data above 511 keV do not
place tight 
constraints on the initial energy of positrons. As discussed by
\citet{2006PhRvL..97g1102B,2006PhRvD..74f3514S}, tighter constraints
come from COMPTEL data combined with the SPI measurements of the 511
keV line flux and $f_{\rm ps}$, restricting $E_{\rm kin}$ to less than 
3--7.5 MeV, depending on the ionization state of the medium \citep[see
  also][for similar calculations based on earlier measurements of the
  gamma-ray flux from the GC region]{1981SvAL....7..395A}.

The limits on the Bulge flux above 511 keV do not strongly
  constrain the amount of cosmic rays in the Bulge, which may produce
  positrons via $\pi^+$ production. Indeed, the same cosmic rays would
  also produce a comparable amount of $\pi^0$, which would be visible
  as gamma-ray emission with a peak around 100 MeV. As discussed by e.g.
  \citet[][]{2000A&A...362..937A}, the total gamma-ray flux from the
  Inner Galaxy 
  associated with $\pi^0$ decay does not exceed $\sim 10^{-4}~{\rm
    phot~cm^{-2}~s^{-1}~sr^{-1}}$, corresponding to a flux  $\lesssim
  10^{-5}~{\rm phot~cm^{-2}~s^{-1}}$ withing the Bulge area (assuming
  a solid angle of the Bulge of $\sim 0.1~{\rm sr}$.  This is about two
  orders of magnitude smaller
  than the observed Bulge flux $10^{-3}~{\rm phot~s^{-1}}$ in the
  annihilation line. The curves shown in Fig.~\ref{fig:ia} are scaled
  by the observed rate of positron production and they approximately match
  the SPI upper limits (for the initial energy of positrons $\sim$100
  MeV). Therefore, the SPI upper limits on the in-flight annihilation of
  positrons constrain the $\pi^+$ production rate to be less than
  $\sim 10^{-3}~{\rm s^{-1}}$. Since the $\pi^+$ and $\pi^0$
  production rates by the same cosmic rays are comparable, it is
  obvious that observations of  $\sim 100$ MeV gamma-rays provide
  much tighter constraints on the amount of cosmic rays in the Bulge.

\subsection{Annihilation in a static ISM}
We now proceed with spectral fitting of the 511 keV line and ortho-positronium
continuum below 511 keV. The simplest possible model is a combination
of a Gaussian at 511 keV to describe two-photon annihilation and the
three-photon spectrum of \citet{1949PhRv...75.1696O}. 
The line
normalization, energy and width and the normalization of the
ortho-positronium continuum are free parameters of the model. The
best-fitting values of these parameters are given in Table
\ref{tab:spec}. The values agree well with the previously reported
SPI results \citep[e.g.][]{2005MNRAS.357.1377C,2006A&A...445..579J}. 
Minor variations of the parameters among different publications can be 
attributed to the different background models and differences in 
the spectra extraction procedures.

Adding a power law component does not improve the $\chi^2$ or
  affect the parameters of the model when the data in the 400--600 keV
  band are considered. As mentioned already at the beginning of
  Section \ref{sec:spec}, this is true when the Bulge
  component of the Bulge+Disk model is considered. It seems that the
  contribution of known compact sources is relatively small around 500
  keV, while the hard continuum associated with cosmic-rays 
  interactions with the gas largely goes into the Disk component.

\begin{table}
\caption{Best-fitting parameters of the Bulge component spectrum (for
  the Bulge+Disk model) in the 400--600 keV band. The quoted errors are
  1$\sigma$ errors for a single parameter of interest. For the
  best-fitting model, $\chi^2=401.8$ for 395 degrees of freedom.
\label{tab:spec}
}
\begin{tabular}{ll}
\hline 
Parameter  & Value \\
\hline
Energy, keV & $510.985\pm0.049$ \\
Width (FWHM), keV & $2.40\pm0.16$ \\
Flux in 511 keV line & $(0.84\pm0.03)~10^{-3}$ \\
Positronium fraction, $f_{\rm ps}$ & $1.00\pm0.02$ \\
\hline
\end{tabular}
\end{table}

\begin{figure*}
\plottwo{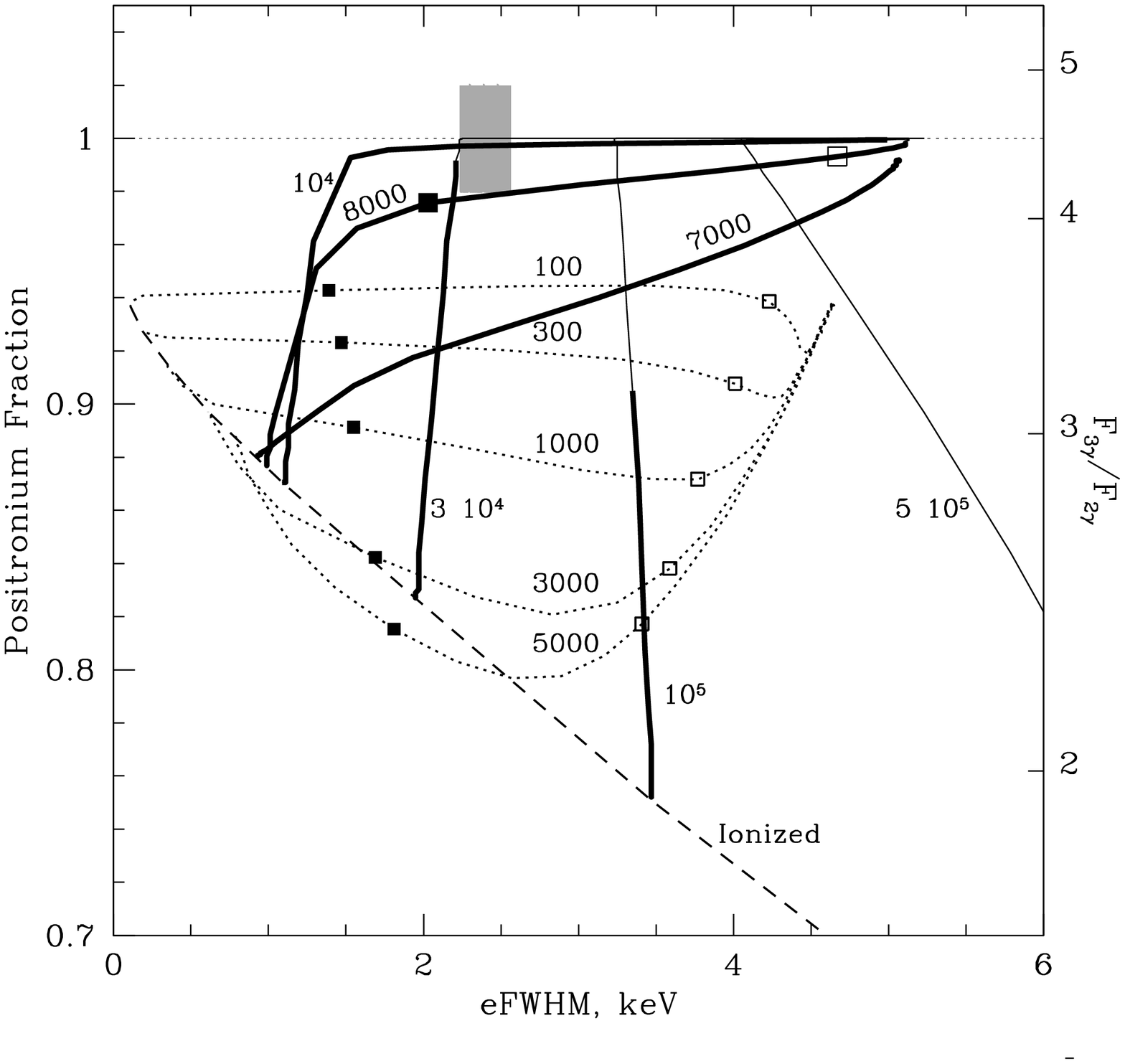}{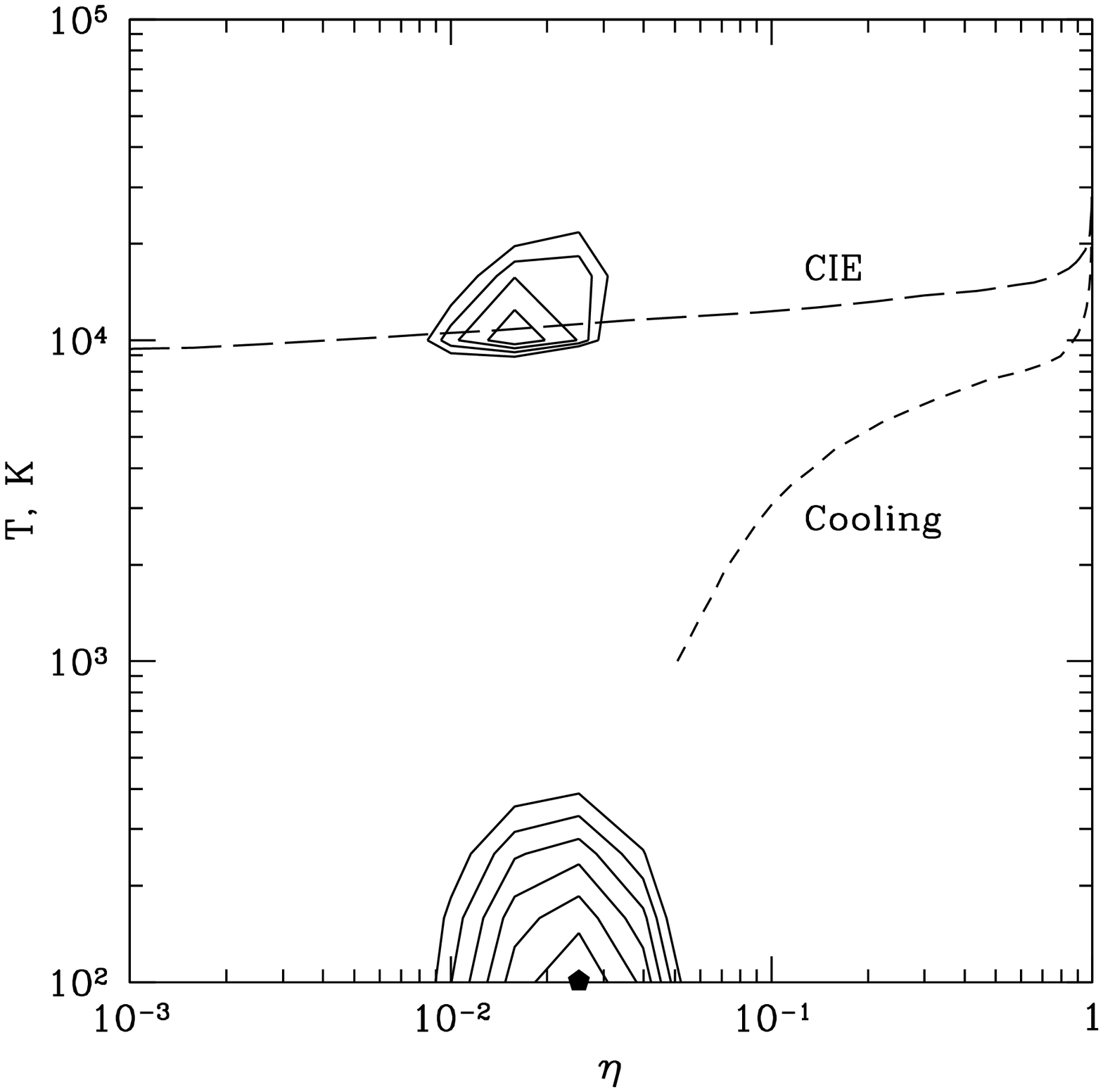}
\caption{{\bf Left:} The effective FWHM of the 511 keV line versus 
  the fraction of annihilations through 
  positronium formation \citep[the curves are adopted
  from][]{2005MNRAS.357.1377C}. The gray  area is the width 
  and the positronium fraction measured by SPI. There are two groups of
  theoretical curves: cold -- $T\le 5000$ K (dotted lines) and
  warm/hot -- $T\ge 7000$ K (solid lines). The temperature is fixed
  for each curve (the labels next to the curves), but the ionization
  fraction varies so that plasma changes from neutral to completely
  ionized along the curve. For cool temperature curves and for the
  8000 K curve, the points corresponding to the ionization degree of
  0.01 and 0.1 are marked with the open and solid squares
  respectively. Each high temperature curve has two regimes, shown by
  thin and thick solid lines, respectively. Thin (thick) lines
  correspond to ionization fractions smaller (larger) than
  expected for collision dominated plasma at this temperature. 
  The over-ionized state (the regime shown by thick lines) is more natural
  for the ISM than the under-ionized one (thin lines). Finally, the
  dashed line shows 
  the relation between the line width and positronium fraction for a
  completely ionized plasma as a function of temperature. {\bf Right:} 
  Countours of $\chi^2$ for a grid of models over the static ISM
  temperature $T$ and ionization degree
  $\eta\equiv n_e/(n_e+n_H)$. The initial energy of positrons is
  set to 500 keV. The $\chi^2$ is calculated for the 400--600 keV range
  for the spectrum of the Bulge component in the Bulge+Disk model. The
  minimum of $\chi^2=402.4$ achieved at $T=10^2$ K and $\eta=0.025$ is marked
  with a black dot. Contours are plotted with an increment $\Delta
  \chi^2=3$. There are two pronounced minima: at $T\sim 100$ and 
$\sim 10^4$ K, both with the ionization degree $\eta \sim$ few percent. The 
``high-temperature'' island has a minimum $\chi^2$ larger than the 
``low-temperature'' solution by $\Delta \chi^2\sim 6$. This difference while
formally statistically significant is reduced when another background
model is used. Long- and short-dashed lines show the relation between
the temperature and the degree of ionization for collisional
ionization equilibrium and radiatively cooling plasma 
(see \S\ref{sec:cooling}), respectively.
\label{fig:ism_fixed}
}
\end{figure*}

As discussed by \citet{1979ApJ...228..928B} \citep[see
  also][]{1991ApJ...378..170G,2005A&A...436..171G}, the effective line
width and the positronium fraction principally depend on the
temperature and the ionization state of the medium. This dependence is
shown in the left panel in Fig.~\ref{fig:ism_fixed} \citep[adopted
  from][]{2005MNRAS.357.1377C}, where the relation between the
effective width of the line and the fraction of positrons annihilating
via positronium formation is shown together with the observed
parameters taken from Table \ref{tab:spec}. Compared to the previously
reported version of the same plot \citep{2005MNRAS.357.1377C}, the
uncertainties on the positronium fraction have decreased
substantially, but the principal result remains the same: in terms of a
single-phase ISM model, a combination of the effective
  width\footnote{The effective width eFWHM \citep{1991ApJ...378..170G}
  of the line is defined as an energy interval containing 
76\% of the line photons.} of the line of $\sim 2.4$
keV and the positronium fraction $f_{\rm ps}\sim 1$ can be explained either
with a gas of temperature in the range from 7500 to $4~10^4$ K or with
  a much cooler gas at $\sim$100 K. In both cases, the ionization
  degree should be at the level of a few \%. While formally the higher
  temperature solution seems to be 
preferable because of the large positronium fraction, our experiments
with background (and spatial) models have shown that we can
drive the positronium fraction down to 96--97\%, making the low-temperature
solution more consistent with the data. 
 
Of course, the shape of the annihilation spectrum in an ISM of
given temperature and ionization state cannot be fully described
by a combination of a Gaussian and the ortho-positronium
continuum. Indeed, in-flight annihilation typically
contributes a broader component to the 511 keV line than the two-photon
annihilation of thermalized positrons. We therefore ran a grid of
models for a set of ISM temperatures and ionization states, calculated
the expected annihilation spectra using the same procedure as in
\citet{2005MNRAS.357.1377C} and compared them with the observed Bulge
spectrum. The initial energy of positrons was set to $E_{\rm kin}=$500 keV. 

The results are insensitive to the particular value of initial energy 
unless $E_{\rm kin}\gg m_ec^2$ or instead $E_{\rm kin}$ is much lower 
than a fraction of keV.
Contours of $\chi^2$ values with an increment $\Delta \chi^2=3$
are shown in the right panel of Fig.~\ref{fig:ism_fixed} .
The minimum of $\chi^2$ achieved at $T=10^2$ K and the ionization degree 
$\displaystyle\eta\equiv n_e/(n_e+n_H)=0.025$ is marked with a 
black dot. As expected, there are two pronounced minima: at $\sim 100$
K and at $\sim 10^4$ K, both at $\eta\sim$ a few per cent. The corresponding
spectra are shown in Fig.~\ref{fig:ism_fixed_spec}; they are further 
discussed in \S\ref{sec:cooling} below. 
In this analysis, the ``high-temperature'' island has a $\chi^2$ larger than 
the ``low-temperature'' solution by $\Delta \chi^2\sim 6$. Given the
simplicity of the 
model, this difference in not sufficient to make a robust conclusion
in favour of the low-temperature solution. We view both solutions as
acceptable. The intermediate values can be reliably excluded. This
non-monotonic behaviour over temperature arises because at 
temperatures in the range 1000--7000 K, i) the typical energy of
a thermalized positron is below the 6.8 eV threshold of positronium
formation via charge exchange with a hydrogen atom and ii) radiative 
recombination with free electrons is suppressed relative to the direct 
annihilation with bound electrons\footnote{For an ionization degree of 
a few \%.}. As a result, the positronium fraction is small 
at intermediate temperatures.
 
\begin{figure*}
\plottwo{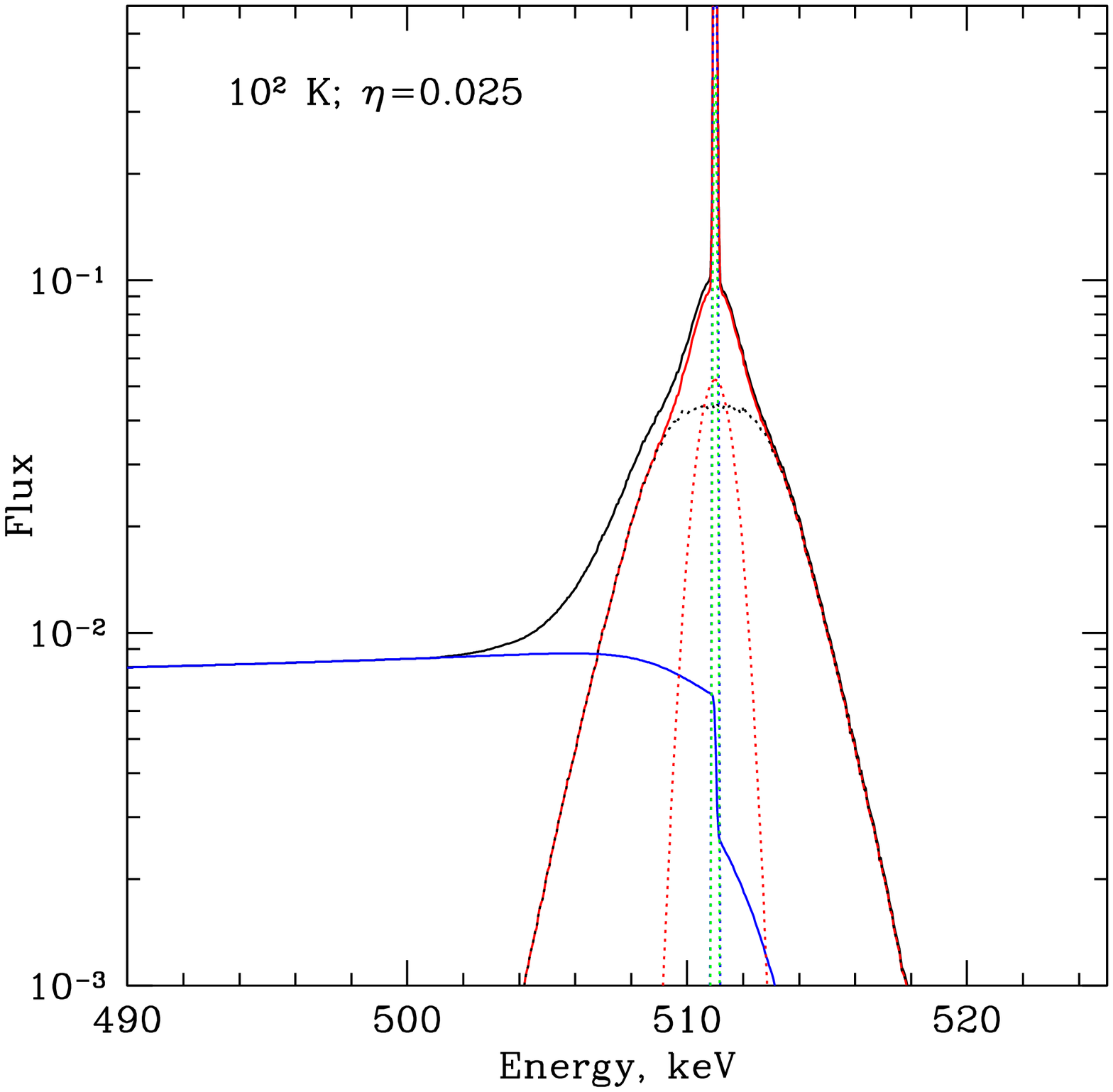}{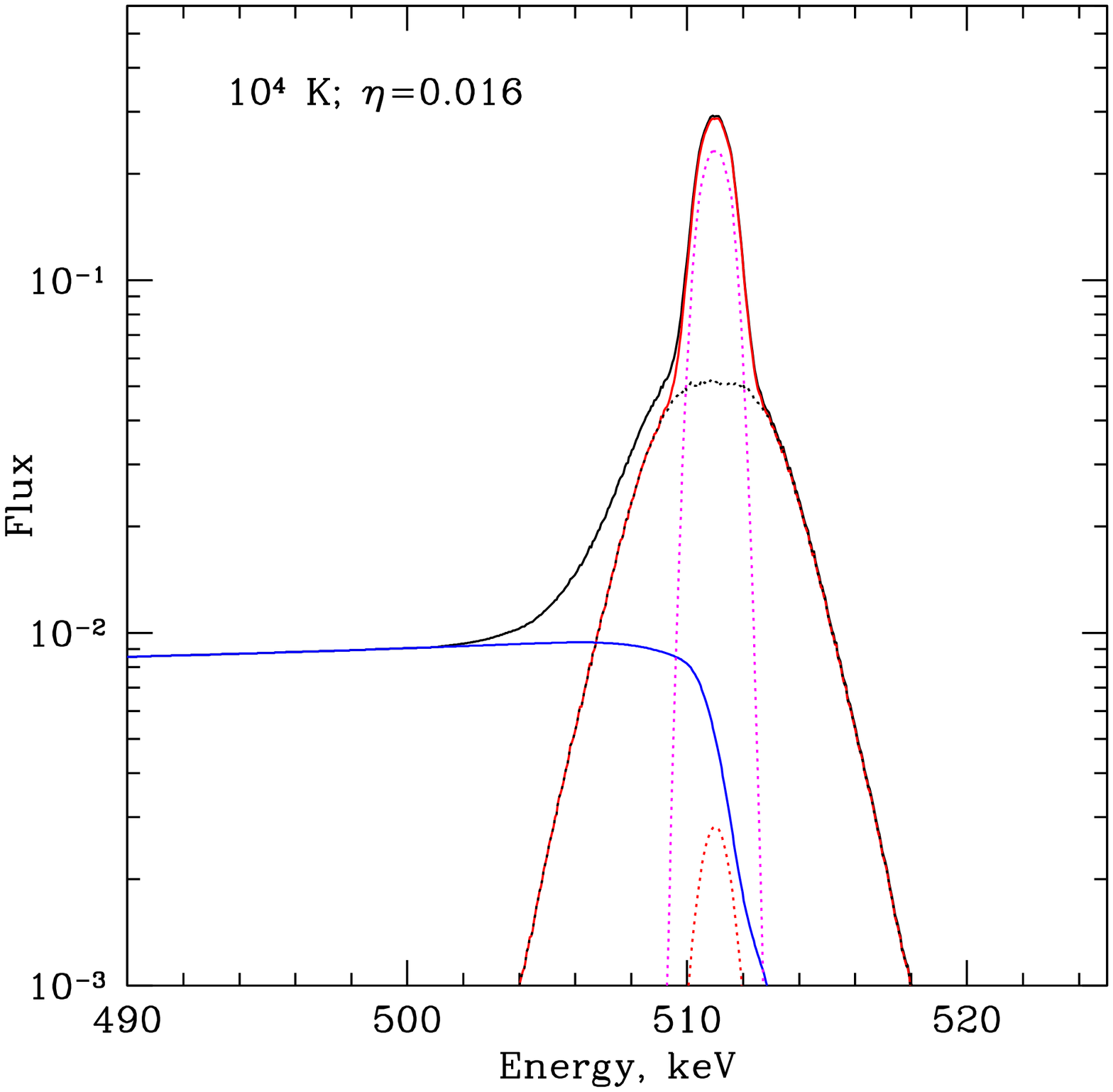}
\caption{{\bf Left:} Expected annihilation spectrum for a 500 keV
  positron in a 100 K plasma with the hydrogen degree of ionization of
  0.025. {\bf Right:} The same as in the left figure, but for $10^4$
  K plasma and $\eta=0.016$. The lines show: solid black line -- total
  annihilation 
  spectrum; solid red line -- total line spectrum; solid blue line --
  othro-positronium continuum; dotted black line -- in-flight charge
  exchange; dotted red line -- thermalized positron annihilation with
  bound electrons, dotted blue line -- radiative recombination of
  thermalized positrons, dotted green line -- direct annihilation;
  magenta -- charge exchange for thermalized positrons.
\label{fig:ism_fixed_spec}
}
\end{figure*}

Overall, the results of this analysis and the implications for the
properties of the ISM broadly agree with previous analyses of SPI data
\citep{2005MNRAS.357.1377C,2006A&A...445..579J} and earlier analyses
by \citet{1979ApJ...228..928B,1991ApJ...378..170G}. 

\subsection{Annihilation in a cooling ISM}

\label{sec:cooling}

\begin{figure}
\plotone{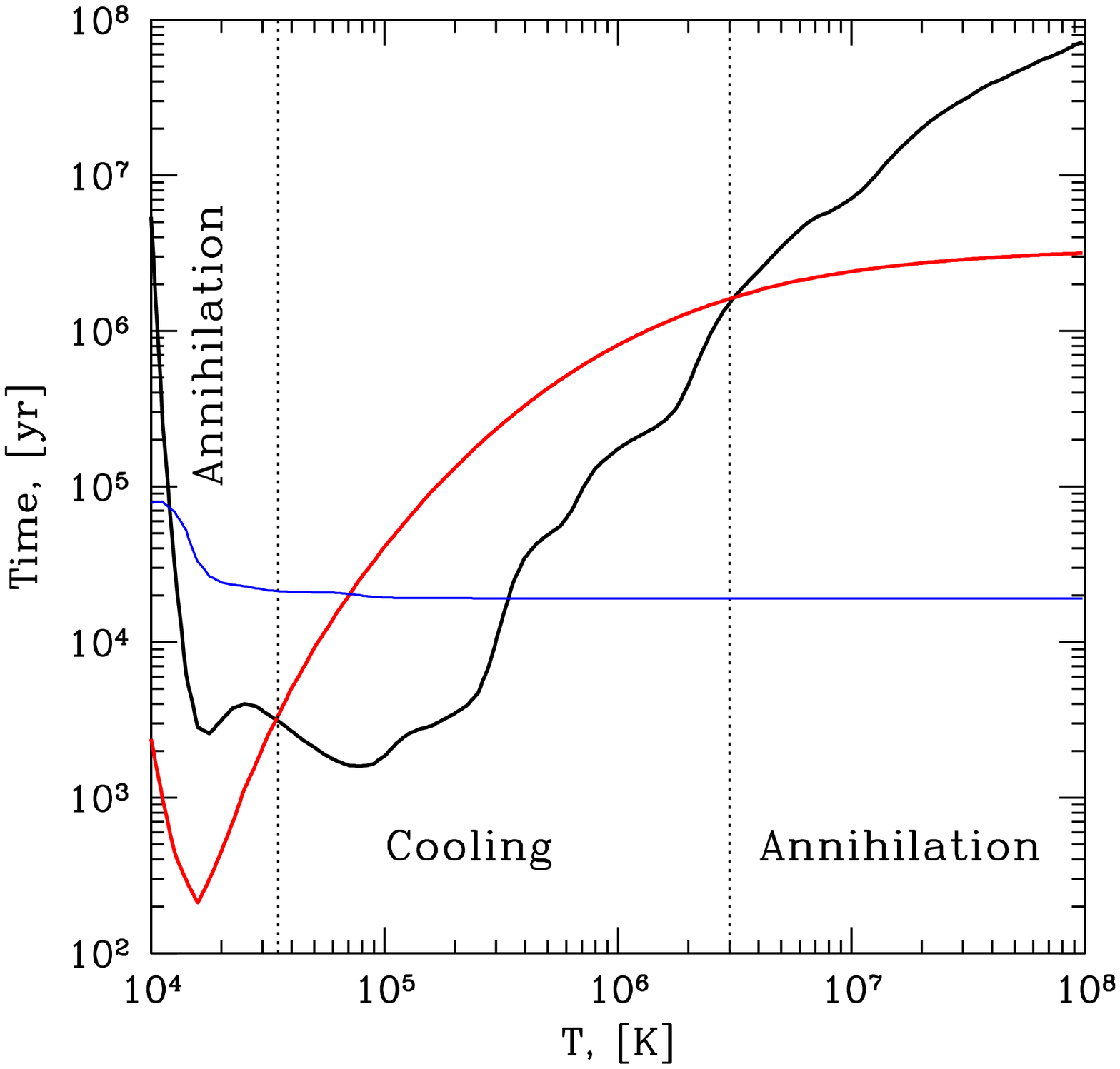}
\caption{Comparison of different time scales relevant for annihilation of
  positrons in a cooling ISM as a function of its temperature: cooling
  of gas (black), slowing down of positrons (blue), annihilation of
  thermalized positrons (red). Calculations are done for the
    proton density $1~{\rm cm^{-3}}$.
The initial energy of positrons is
  assumed to be 500 keV. The dotted vertical lines separate the areas
  where cooling or annihilation dominates. For temperatures in the
  range from $\sim 3~10^4$ to $\sim 3~10^6$~K, the gas can cool
  faster than thermalized positrons annihilate. For instance, if 500
  keV positrons are born in a $\sim 10^6$~K ISM, the gas will cool down
  to a few $10^4$~K before the positrons have a chance to
  annihilate. Once the temperature has dropped, the thermalized
  positrons will annihilate and the observer will see a gamma-ray spectrum
  indicating annihilation in a $\sim 10^4$~K gas, regardless of the
  initial ISM temperature.
\label{fig:tcool}
}
\end{figure}

The model spectra used in the previous section were calculated assuming
that the temperature and the degree of ionization of the ISM do not
change on the time scales of positron slow-down and annihilation. This
is not an obvious assumption since the time scale  
for the annihilation of thermalized positrons can be longer than the
ISM radiative cooling time scale, as shown in Fig.~\ref{fig:tcool}. We
define an isobaric cooling time scale (black line in
Fig.~\ref{fig:tcool}) as the ratio of the gas enthalpy to the
radiative losses: 
\begin{eqnarray}
t_{\rm cool}=\frac{\gamma}{\gamma-1}\frac{nkT}{n^2\Lambda(T)}, 
\label{eq:tcool} 
\end{eqnarray}  
where $\gamma=5/3$ is the gas adiabatic index, $n$ the total gas
density, $k$ the Boltzmann constant, $T$ the gas temperature, and $\Lambda(T)$
is the appropriately normalized cooling function, adopted from
\citet{1993ApJS...88..253S}. The slow-down time estimated for 500 keV
positrons is shown with the blue line; for this estimate we 
considered only Coulomb and hydrogen ionization losses (ignoring excitation
losses and the contribution of helium). The temperature dependence of the
slow-down time scale in Fig.~\ref{fig:tcool} arises due to the variations
of the degree of ionization with temperature. For this plot we adopted
the collisional ionization equilibrium from \citet{1985A&AS...60..425A}.
Finally, the red line shows the time scale for annihilation of
thermalized positrons, which includes direct annihilation with free and
bound electrons, charge exchange with hydrogen and radiative recombination.

From Fig.~\ref{fig:tcool} it is clear that for temperatures in the
range from $\sim 3~10^4$ to $\sim 3~10^6$~K, the gas is able to cool
faster than thermalized positrons can annihilate. For temperatures
outside this range, the annihilation time is shorter than the cooling
time and the medium can to a first approximation be regarded as
static. The most interesting conclusion that can be drawn from the
comparison of time scales in Fig.~\ref{fig:tcool} is that if $\sim
500$~keV positrons are born in a $\sim 10^6$~K ISM, then the gas
temperature will drop\footnote{In the absence of an external source of
  heat.} to a few $10^4$~K before the positrons have a chance to
annihilate. Once the ISM has cooled down, the positrons will
annihilate and the observer will measure spectral characteristics
typical of annihilation in warm gas, with no indication
that the positrons were born in a hot medium.

We now proceed with calculating spectra expected for annihilation of
positrons in a cooling ISM. This implies solving a time dependent
problem: for given initial gas temperature $T_0$ and kinetic energy of
positrons $E_0$ find the annihilation spectrum emitted while the gas
cools down to a given final temperature $T_1$. We solve 
this problem using a modified version of our Monte Carlo code, which retains 
from the original version the algorithms for description
of slowing down, thermalization and annihilation of positrons (in
a pure hydrogen plasma).

The outcome of this calculation crucially depends on the assumptions
made about the ionization state of the ISM during its cooling from
$\gtrsim 10^6$ to $\lesssim 10^4$~K. In real astrophysical 
situations, a radiatively cooling plasma is expected to be
substantially overionized relative to collisional ionization
equilibrium (CIE), since the characteristic cooling time is shorter than the
time scale of hydrogen recombination. This
will influence both the ISM cooling rate and the fate of
positrons. We therefore adopt the temperature dependencies for
the non-equilibrium ionization state and cooling rate of an
isochorically cooling gas from \cite{1993A&A...273..318S} and
\cite{2007ApJS..168..213G}, which cover the temperature range from
$10^3$~K to $10^9$~K (we assume that hydrogen is fully ionized at
$T>5~10^6$~K). The temperature dependencies of the hydrogen
ionization fraction for the CIE and non-CIE cases are compared in the
right panel of Fig.~\ref{fig:ism_fixed}. 

\begin{figure}
\plotone{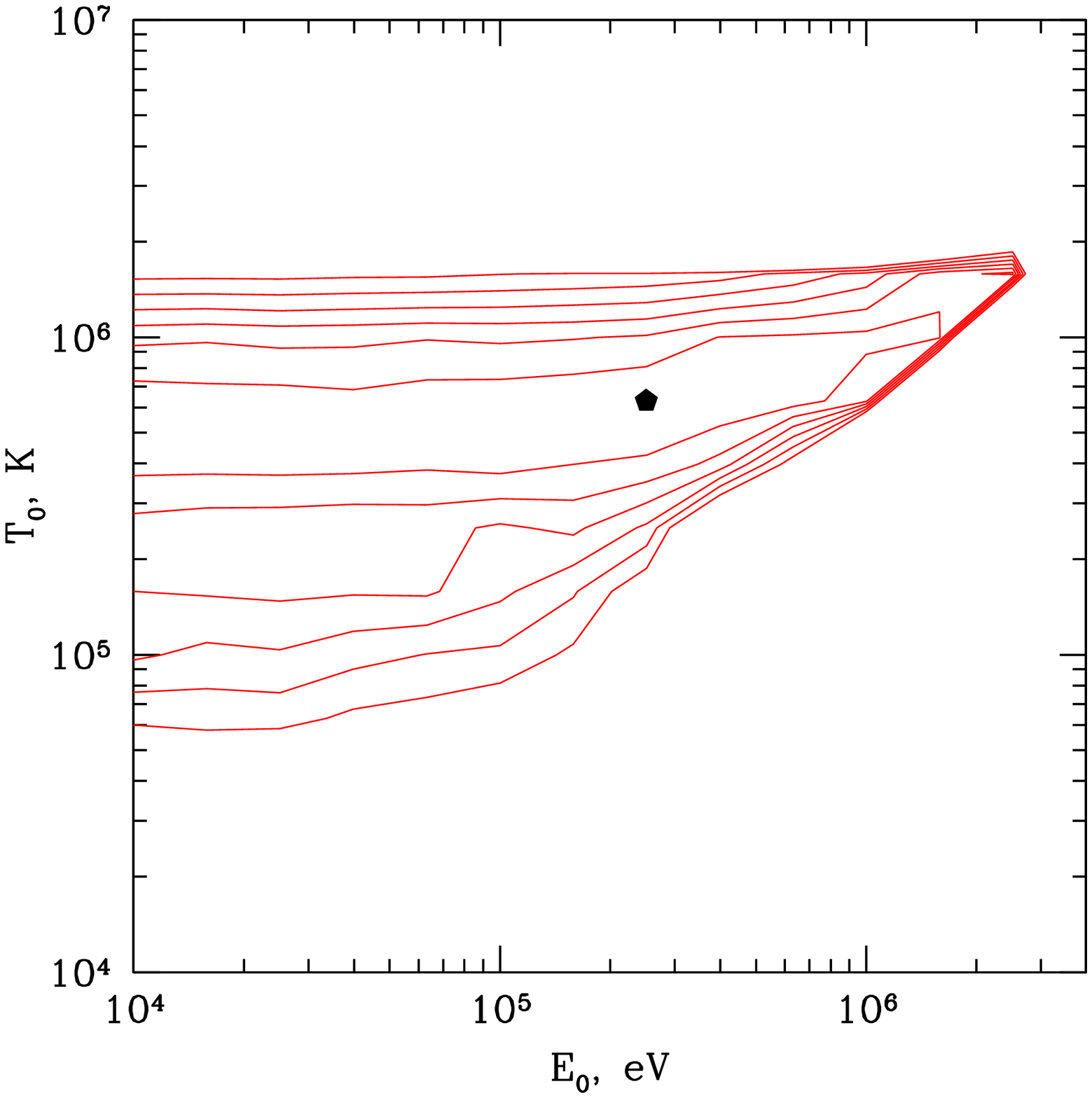}
\caption{Countours of $\chi^2$ for a grid of models for a cooling ISM
  over the range of initial ISM temperatures $T_0$ and initial positron energy 
  $E_0$. The minimum of $\chi^2=400.8$ achieved at $T_0=6.3~10^5$ K 
  and $E_0=250$ keV is marked with a black dot. Countours are plotted with an increment $\Delta
  \chi^2=3$. 
\label{fig:ism_cooling}
}
\end{figure}

We carried out simulations covering a broad range of 
parameter values: $E_0=$10~keV--10~MeV, $T_0=10^4$--$10^7$~K. 
A run was terminated when
the ISM cooled down to $10^3$~K; it turns out that the absolute
majority of positrons annihilate before the gas reaches this
temperature, except for extreme values of the initial ISM temperature
and positron energy. Although we assumed isochoric cooling,  
this assumption is not crucial, since the case of isobaric cooling is 
characterized by very similar ISM ionization state and cooling rate
\citep{1993A&A...273..318S}. 
 
\begin{figure*}
\plottwo{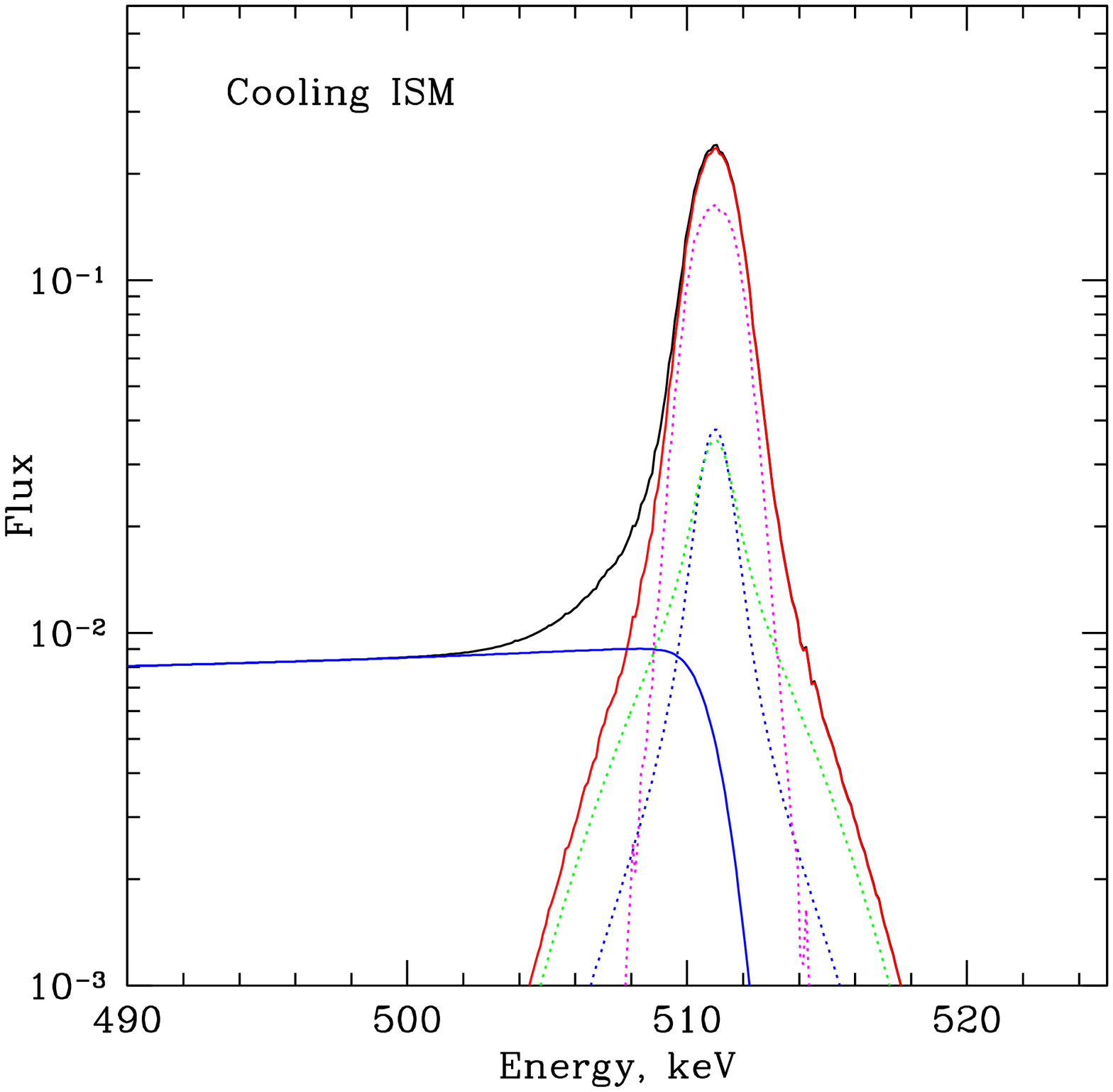}{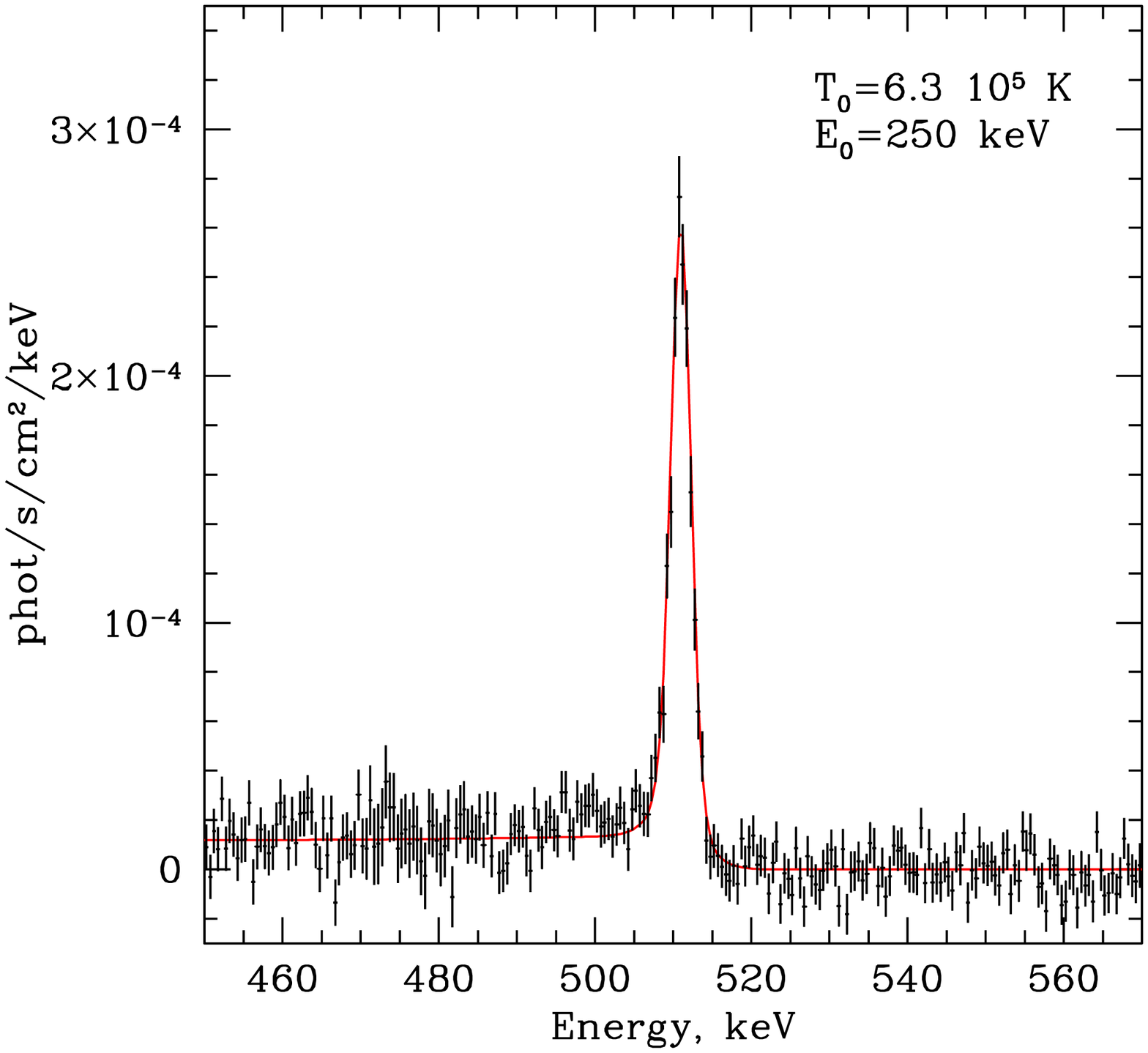}
\caption{{\bf Left:} Expected annihilation spectrum for a radiatively cooling plasma with 
  initial temperature $T_0=6.3~10^5$ K and initial positron kinetic energy 
  $E_0=250$ keV. Different lines show different spectral components, 
  as in Fig.~\ref{fig:ism_fixed_spec}. In contrast to the annihilation in
  a static medium,
in-flight component makes negligible contribution for the cooling ISM solution, in which case the
annihilation spectrum is essentially a superposition of spectra generated
by thermalized positrons in gases with temperatures ranging from 
$\sim 10^4$ to $\sim 2~10^5$~K.
 {\bf Right:} Spectrum of the annihilation radiation for a cooling ISM with
  initial temperature $T_0=6.3~10^5$ K and initial positron energy
  $E_0=250$ keV in comparison with the observed Bulge component spectrum in the
  Bulge+Disk model.
\label{fig:ism_cooling_spec}
\label{fig:spec_cooling_best}
}
\end{figure*}

Similarly to the case of annihilation in a static medium considered
in the previous section, we fit the spectra predicted by our
cooling ISM models to the spectrum measured by SPI. In
Fig.~\ref{fig:ism_cooling}, we show contours of  
$\chi^2$ values on the ($E_0$, $T_0$) plane. The best-fitting parameters are
$E_0\sim 250$~keV and $T_0\sim 6~10^{5}$~K (see the corresponding
model spectrum in Fig.~\ref{fig:ism_cooling_spec} and its fit to the SPI data 
in Fig.~\ref{fig:spec_cooling_best}), with a large associated uncertainty 
region: the initial gas temperature and positron energy are allowed to
take values from $\sim 2~10^5$ to $\sim 10^6$~K and from 10~keV 
(the lower boundary of our simulations) to $\sim 2$~MeV, respectively.  
Therefore, the predicted gamma-ray spectrum is fairly insensitive 
to the initial parameters of the ISM and positrons. The reason for
this is that in a broad range of $E_0$ and $T_0$ values the main
contribution to the total spectrum is provided by annihilations of
thermalized positrons during the gas cooling from $\sim 10^5$~K to
$\sim 10^4$~K (when the positron annihilation time is a few times the ISM
cooling time, see Fig.~\ref{fig:tcool}).

Judging by the small differences between the minimum $\chi^2$ values, 
summarized in Table \ref{tab:ism}
(and recalling again that there are non-negligible systematic
uncertainties associated with the data analysis), our best-fitting static
ISM and cooling ISM models describe the 
SPI spectrum nearly equally well. In other words, we cannot strongly
favour any of these models based on the available spectral
data. In reality, this is mostly owing to the finite energy
resolution of SPI. This becomes clear if we compare 
the spectra predicted by the different models, shown in  
Figs.~\ref{fig:ism_fixed_spec} and \ref{fig:ism_cooling_spec}.

\begin{table}
\caption{Values of $\chi^2$ for various spectral models for the Bulge
  component  spectrum in the 400--600 keV band (400 channels, 0.5 keV wide). 
\label{tab:ism}
}
\begin{tabular}{lll}
\hline 
Model &Parameters & $\chi^2$  \\
\hline
Static ISM & $T=10^2$ K; $\eta=0.025$ & 402.4\\ 
Static ISM & $T=10^4$ K; $\eta=0.016$ & 408.6\\ 
Cooling ISM & $T_0=6.3~10^5$ K; $E_0=250$ keV & 400.8\\ 
\hline
\end{tabular}
\end{table}

\section{Discussion}

A comprehensive answer to the fundamental question of the origin of
the bulk of annihilating positrons should explain a number of 
key observational facts, including the total production rate of
positrons ($\sim 10^{43}$ positrons per second within the Galactic 
bulge), the dominance of the bulge compared to the disk,
the morphology of the 511 keV line emission (a 8\deg $\times$ 6\deg
"Bulge" plus a more extended, along the Galactic plane,
``Disk'' component) and the spectral properties 
of the Bulge component, which (within the accuracy allowed by the SPI
energy resolution) resemble the annihilation of positrons in a warm
($\sim 10^4$~K) or cold ($\sim 10^2$~K) medium with the ionization
degree of a few per cent.

On these grounds (flux and morphology), such straightforward
explanation as $^{26}$Al or the interaction of cosmic rays with the ISM
are disfavored. The morphology problem can be partly alleviated
by assuming transport of the positrons away from the place where
  they are born \citep[e.g.][]{2006A&A...449..869P}. Recently, 
an attempt to provide a self-consistent model was done
by \citet{2009ApJ...698..350H,2009PhRvL.103c1301L}. Their model
assumes that supernovae are the source of positrons (via $\beta^+$ decay
  of $^{56}$Ni, $^{44}$Ti and $^{26}$Al) in the Galaxy
and the propagation of positrons in the ISM is different in the bulge
and in the disk. The difference in propagation explains the weakness of
the disk compared to the bulge -- a large fraction of positrons born in
the disk escape into the halo, while only a small fraction of positrons
escape the bulge. In the bulge, the positrons diffuse through the ISM
and annihilate when entering the HII and HI envelopes of molecular
clouds -- this explains the spectral properties of the annihilation
emission. Finally, the morphology of the bulge is explained by 
annihilation in a Tilted Disk within 1.5 kpc of the GC. 
The key assumptions in the Higdon et al. model are that i)
  magnetic flux tubes are almost radial in the Bulge region and ii)
  positrons propagation along the flux tubes can be treated in
  the diffusion approximation with the frequency of scattering (or the
  mean free path) estimated from scaling the energy density of
  magnetic fields down to the Larmor radius. Their estimate of the
  mean free path in the Bulge region is large enough to allow
  positrons to reach the Tilted Disk (see \S\ref{sec:tilt}) 
  before they slow down and annihilate. Thus the morphology of the
    Tilted disk gives the source of annihilation radiation in the GC
    its characteristic shape.

Here we consider the alternative possibility that the thermal
  evolution of the ISM plays the key role in the GC annihilation
  emission. The key assumption of the model is that the
  medium in which the bulk of positrons are produced and annihilate
  has a cooling time shorter than the slow down and annihilation time
  scales. From Fig. \ref{fig:tcool} it follows that the temperature
  ought to be lower than $\sim 3~10^6$K. 

 Gas with temperature in the range $10^5$-- few $10^6$~K
  is observed in the Galaxy, as indicated by e.g. CVI, OVII and OVIII
  lines \citep[e.g.][]{2002ApJ...576..188M}. An ISM with temperature
  $\sim 3~10^6$ K is also observed in the central region 
  of M31 \citep{2008MNRAS.388...56B}. The origin of this gas is not
  clear. For a long time observations of 6.7 and 6.9 keV lines of
  heavily ionized iron and an unresolved X-ray glow (at energies above
  a few keV) of the Galactic plane were regarded as evidence for a much
  hotter, $\sim 10^8$ K, plasma filling most of the bulge's volume
  \citep[e.g.][]{1986PASJ...38..121K}. Positrons injected into such a
  medium will slow down and annihilate before the plasma is able to
  cool (Fig. \ref{fig:tcool}), and a very broad annihilation line
  without an ortho-positronium continuum is expected to be observed, in
  stark contrast with observations. However, recent observations
  \citep{2006A&A...452..169R,2006A&A...450..117S,2009Natur.458.1142R}
  have convincingly demonstrated that the apparently diffuse Galactic X-ray
  emission is in fact a superposition of millions of faint compact sources --
  accreting white dwarfs and coronally active stars. This removes the
  problem of confining $\sim 10^8$ K gas by the gravitational well of
  the Galaxy. Cooler, $10^5$-- few $10^6$~K, gas can on the contrary
  be trapped in the Galaxy potential. For instance, the density of 
  $2~10^6$K gas in hydrostatic equilibrium in the Milky Way potential
  would drop by a factor of $\sim 3~10^{3}$ at a distance of $\sim
  1.5$ kpc above the plane. This means that such gas can accumulate in
  the Galactic bulge. 

  However, a self-consistent model explaining the
  few $10^6$~K ISM in the bulge of the Milky Way or in M31 is still
  missing. If one takes the mass loss rate by evolving stars
  \citep[e.g.][]{1991ApJ...376..380C} and the total energy and iron
  input rate from SNIa \citep[e.g.][]{2008MNRAS.383.1121M}, then a
  ``natural'' temperature of the gas is of order few keV (mean energy
  per injected particle) and the expected abundance of iron is about 5
  times the Solar value. These parameters do not fit the properties of
  the observed few $10^6$ K component of the
  ISM. \citet{2009MNRAS.398.1468T} and \citet{2010MNRAS.tmp.1115T} 
  argue that the account for non-uniformity and intermittence in
  energy injection by SNIa leads to a broad gas distribution over
  temperature with a lower than the mean temperature component making
  largest contribution to the observed X-ray emission. It is not yet
  clear if this provides a full explanation of the origin of $10^6$ K
  gas. We therefore make an  ad hoc assumption that a few million
  degree medium is widespread in the Milky Way bulge.

The positron annihilation in a cooling ISM (\S\ref{sec:cooling}) can 
 resolve some of the issues mentioned at the beginning of
this section. Indeed, if the ISM is able to cool down to $10^5$--$10^4$~K
before the positrons annihilate, then the observer will always see
a spectrum characteristic of annihilation in a warm plasma. This resolves 
the question of why positrons are apparently annihilating in a
warm/cold medium even though such a medium should have a relatively
small filling factor. Otherwise, the positrons would have to be
produced preferentially in the warm and/or cold ISM phases or to
migrate between phases. In the cooling ISM model this is not
required. If the ISM is dynamic and goes through cooling/heating
cycles, then even those positrons born in the hot phase
will annihilate in a warm ($10^4$--$10^5$~K) gas. Transport of positrons
over large distances is not required in this model. Some moderate
transport of positrons will not contradict the model, since it will just
make shorter the lifetimes of positrons initially born in the hot ISM, 
but the positrons will still predominantly annihilate while
migrating through the warm/cold ISM phases.

Putting all arguments together, it seems plausible that the following
model might work:
\begin{itemize}
\item Positrons are injected in volume-weighted fashion into the ISM
  (e.g. by SNIa). This means that most of the positrons are initially
  in the hot ISM phase. The key requirement of the model is that
    the large volume fraction has to be occupied by a few $10^6$ K gas
  -- which may not be true in the Galactic bulge.
\item The plasma cools radiatively before the positrons slow down. An
  outflow, adiabatic expansion of the gas advect/move positrons and
  may contribute to positron cooling.
\item Eventually, the plasma reaches relatively low temperatures and
  the positrons annihilate, producing the spectral features as observed by SPI.
\end{itemize}

If correct, this model predicts that there should be some similarity
in the spatial distribution of the annihilation emission and the
thermal emission of plasma with temperature in the $10^4$--2~$10^5$~K range. Outflows from the central regions of 
normal galaxies are well-known. A good example is M31 with clear evidence of hot
$\sim 3~10^6$ K gas \citep{2008MNRAS.388...56B}, possibly outflowing
from the central region. In subsequent publications we plan to consider the 
dynamics of galactic outflows, advection and annihilation of positrons and the 
expected spectral signatures (in soft X-ray and other wavebands) of the main 
annihilation sites.

\section{Conclusions}
We have analysed SPI/{\it INTEGRAL} data on the 511 keV line from the
Galactic Centre, accumulated over $\sim$6 years of observations.

Using a simple two-component Bulge+Disk model we have constructed the
broad-band energy spectra of both components. The Bulge component
shows a prominent 511 keV line and essentially no flux at 1.8 MeV,
while the Disk component on the contrary contains a prominent 1.8 MeV
line and a very weak annihilation line. The inferred annihilation
rate in the Bulge component is $\sim 10^{43}$ positrons per second
(for $f_{\rm ps}=1$).

The FWHM of the 511 keV in the Bulge is $2.40\pm 0.16$ keV. The flux
ratio of the line and three-photon continuum corresponds to a 
fraction of positrons annihilating via positronium formation close to
unity ($f_{\rm ps}=1.00\pm 0.02$). Our experiments with using slightly
different SPI background models have yielded $f_{\rm ps}$ in the range
0.96--1. This combination of spectral parameters can be explained
in terms of a single-phase static ISM model, namely as annihilation of
positrons in a warm ($\sim 10^4$~K) or cold ($\sim 10^2$~K) medium with
the hydrogen ionization fraction of a few per cent.

We have demonstrated that the observed annihilation spectrum of the
Bulge component can also be explained with a model of positrons
annihilating in a radiatively cooling ISM, provided that the initial
temperature of the ISM is in the range $2~10^5$--$10^6$~K. In such a
medium, the ISM cools down to $\sim 10^5$--$10^4$~K before the positrons
are able to annihilate and the observer sees an annihilation spectrum
typical of a warm ISM. This suggests that the positrons might be
initially injected into the hot ISM phase, which has the largest filling
factor in the Galaxy, but the annihilation occurs only when the gas
has cooled down to a relatively low temperature, effectively erasing
the memory on the initial conditions (the initial 
temperature and ionization state of the ISM and the initial energy of
positrons). If there is an outflow of hot gas, then the
positrons will be advected with the flow and annihilate in the part of the
flow where the temperature has dropped to $10^5$--$10^4$~K. The mass of the
cooling gas need not to be large -- it is enough to assume that the
initially hot ISM phase occupies the largest fraction of the volume
where the positrons are produced, provided that positrons do not
migrate freely between phases. The key requirement of the model
is that the large fraction of the bulge's volume is occupied by a few
$10^6$ K gas. In this paper this is adopted as an ad hoc assumption.  The
  origin of this gas in not entirely clear, although observations suggest that
  this component is present in the central part of the Galaxy.

\section{Acknowledgements} 
We are grateful to Roland Diehl, Umberto Maio, Sergey Molkov and
Jean-Pierre Roques for useful discussions. This work was supported by
the DFG grant CH389/3-2, grant NHs-5579.2008.2 of the Russian
Foundation for Basic Research and programmes P-07 and OFN-17 of the
Russian Academy of Sciences. SS acknowledges the support of the
Dynasty Foundation.


\label{lastpage}
\end{document}